\documentclass[12pt]{article}
\usepackage[a4paper, margin=2in]{geometry}

\usepackage[utf8]{inputenc}
\usepackage[T1]{fontenc}
\usepackage{soul}

\usepackage{amsmath,amsfonts,amsthm,amssymb}
\usepackage{graphicx}
\usepackage{longtable}
\usepackage{array,multicol}
\usepackage{setspace}
\usepackage{hyperref}
\usepackage{subcaption}  

\usepackage[square,sort,comma,numbers]{natbib}
\usepackage{fancyhdr}
\usepackage[Glenn]{fncychap}
\usepackage{fullpage}
\usepackage{xcolor}
\definecolor{sepia}{rgb}{0.44, 0.26, 0.08}
\def\R{\mathbb{R}}

\def\DD{{\cal D}}

\def\K{{\cal K}}

\let\eps=\varepsilon

\newcommand{\Fb}{{\boldsymbol F}}
\newcommand{\Lb}{{\boldsymbol L}}

\newcommand{\bs}{{\boldsymbol \sigma}}
\newcommand{\bbs}{{\boldsymbol b}}

\newcommand{\bm}{{\boldsymbol m}}
\newcommand{\bu}{{\boldsymbol v}}
\newcommand{\D}{{\boldsymbol D}}
\newcommand{\W}{{\boldsymbol W}}

\newcommand{\Rb}{{\boldsymbol R}}
\newcommand{\M}{{\boldsymbol M}}
\newcommand{\Pb}{{\boldsymbol P}}

\newcommand{\Ub}{{\boldsymbol U}}

\newcommand{\Ib}{{\boldsymbol I}}

\newcommand{\Qb}{{\boldsymbol Q}}

\newcommand{\be}{{\boldsymbol e}}

\renewcommand{\R}{\mathbb{R}}

\newcommand{\bro}{{\boldsymbol \rho}}

\newcommand{\btau}{{\boldsymbol \tau}}
\newcommand{\bTau}{{\boldsymbol {\cal T}}}
\newcommand{\Tau}{{ \cal T}}
\newcommand{\Ac}{{ \cal A}}
\newcommand{\MM}{{ \cal M}}

\numberwithin{equation}{section}

\hypersetup{
  colorlinks   = true,    
  urlcolor     = blue,    
  linkcolor    = blue,    
  citecolor    = blue     
}

\begin{document}
%
\title{Dislocation saturation in slip rate  driven processes  and initial microstructure effects for large plastic deformation of crystals}

\author{
    Jalal Smiri\footnote{LSPM (UPR 3407 of CNRS), Université Sorbonne Paris Nord, 93430, Villetaneuse, France} 
    \footnote{now at  FEMTO-ST, Université Marie et Louis Pasteur, 25000-Besançon, France}, 
    O\u{g}uz Umut Salman$^*$, 
    and Ioan R. Ionescu$^{*}$\footnote{IMAR, Romanian Academy, 10587-Bucharest, Romania} \footnote{corresponding author, ioan.r.ionescu@gmail.com}
}\date{version : \today}
\maketitle

\begin{abstract}
Dislocation-density-based crystal plasticity (CP) models are introduced to account for the microstructural changes throughout the deformation process, enabling more quantitative predictions of the deformation process compared to slip-system resistance-based plasticity models. In this work, we present a stability analysis of slip-rate-driven processes for some established dislocation density-based models, including the Kocks and Mecking (KM) model and its variants. Our analysis can be generalized to any type of dislocation density model, providing a broader framework for understanding the stability of such systems. We point out the existence of saturation dislocation densities {\color{black} and the essential role of initial dislocation density  in distinguishing between hardening and softening responses.}
Since the initial microstructure, modeled through the dislocation density, could be related to the size or the sample preparation process, implicit size-dependent effects can also be inferred.
To further explore these phenomena, we conduct numerical simulations of pillar compression using an Eulerian crystal plasticity framework. Our results show that dislocation-density-based CP models effectively capture microstructural evolution in small-scale materials, offering critical insights for the design of miniaturized mechanical devices and advanced materials in nanotechnology.\end{abstract}


\section {Introduction} 
Crystalline solids undergo plastic deformation when macroscopic stresses exceed critical thresholds. This non-recoverable deformation is primarily mediated by the nucleation, motion, and interaction of dislocations - line defects in the crystal lattice structure~\cite{Devincre2008-to,Cui2020-wk}. These dislocations evolve within a complex energy landscape shaped by both applied stresses and multi-scale interactions, including long-range elastic fields and short-range core effects. Controlling crystal plasticity mechanisms is essential across numerous applications, including metal strengthening processes, forming operations \cite{ref2:habibfracture}, fatigue resistance enhancement \cite{Irastorza-Landa2016-ir,Guo2020-ho,Prastiti2020-qo}, nano-scale manufacturing, and micro-pillar optimization for miniaturized mechanical components \cite{ref4:takata2023influence,Zhanga2016-mx,Pan2019-sl,Weiss2021-pz,Wu2023-he,Wijnen2025-ax}.

This complexity in crystal plasticity spans multiple length scales, creating a hierarchical structure of deformation mechanisms. At the atomistic level, individual dislocation cores and their interactions with grain boundaries govern local plastic events~\cite{Baruffi2019-qw}. These microscopic processes aggregate into mesoscale phenomena, including dislocation patterns, cell structures, and grain-level deformation. Remarkably, despite this underlying complexity, macroscopic plasticity often manifests as a smooth, continuous stress-strain response. This observation has led to the widespread adoption of continuum approaches that model plastic flow as a continuous spatio-temporal deformation mechanism in conventional engineering applications, effectively homogenizing the rich microstructural landscape.

Continuum Crystal Plasticity (CP) incorporates lattice-based kinematics into the classical continuum framework for modeling crystal plasticity. Its original mathematical formulation was introduced by Hill \cite{Hill1966-xr} and Hill and Rice \cite{Hill1972-vl}, with initial applications by Asaro and Rice \cite{Asaro1977-nn,Asaro1983-cw} and Pierce, Asaro, and Needleman \cite{Peirce1983-sm}. Since then, many authors have further developed CP theory (see \cite{Roters2010-fv} for a comprehensive overview). The classical continuum crystal theory of plasticity assumes that crystalline materials undergo irreversible flow when applied stresses exceed certain thresholds. This theory models the stress-strain response with continuous curves, implying that the  heterogeneities are averaged out. This approach has been successful in reproducing key plasticity phenomena such as yield, hardening, and plastic shakedown.

Obviously, CP theory does not resolve individual plastic defects. Instead, dislocations are represented by continuously evolving incremental shear strains (slips) $\gamma^{s}$. This coarse representation conceals both short-range and long-range dislocation interactions. However, the simplified kinematics allows access to larger time and length scales, enabling the modeling of complex 3D systems with intricate geometries. The method can be implemented in a Finite-Element Method (FEM) setup or in spectral methods, facilitating the modeling of various physical phenomena, from grain boundary evolution to pattern formation under cyclic loading \cite{Roters2010-fv,Lamari2025-ro,Lame-Jouybari2024-yt,HARDIE2023103773}.

While traditional CP models use phenomenological constitutive laws, more advanced formulations incorporate dislocation-related effects, such as kink and shear-band formation \cite{ref15:cai2021strain}. These advanced models employ physics-based constitutive laws that rely on internal variables with micro-structural significance.  {\color{black} Among these internal variables, dislocation density stands out in the literature (see for instance \cite{ZHANG2023103553, CAPPOLA2024103863,FISCHER2022144212,HUNTER2022103178,EGHTESAD2023103646}) as the most crucial in the context of plasticity as primary carriers of plastic deformation.}
By incorporating dislocation densities as state variables, CP models can address numerous complex scenarios: interactions between dislocations and grain boundaries~\cite{ref16:hamid2017modeling}, prismatic slip in $\alpha$-titanium alloys~\cite{ref17:ghorbanpour2017crystal}, deformation of single FCC crystals under high strains~\cite{Ye2023-hk}
, size-dependent plastic flow~\cite{ref20:cackett2019spherical}, and precipitate-induced cyclic softening~\cite{Agaram2021-es}, among others.

Building upon these dislocation density-based crystal plasticity frameworks, the objective of this study is to perform a stability analysis of slip rate driven processes for several established models: the Kocks and Mecking (KM) model, Kocks-Mecking-Teodosiu model, Kocks-Mecking-Estrin model, and Cui-Lin-Liu-Zhuang (CLLZ) model. Our analysis aims to identify the conditions leading to the attainment of "saturation" states of dislocation densities in active slip systems. Additionally, we seek to evaluate the (linear) stability of these states to determine whether they demonstrate attractor characteristics, which would provide insight into the long-term (large deformation)  evolution of microstructure under continued loading.

The stability analysis of dislocation densities (DD) provides essential insights into dislocation evolution across different initial microstructures, helping to identify the conditions under which material softening or hardening occurs. This finding is consistent with the conclusions in~\cite{Jennings2010-yg}, which suggest that the initial microstructure controls the response of nanopillars,  influencing the occurrence of size-effects. Notably, the dislocation density may be influenced by specimen size or sample preparation processes (see \cite{10.1063/1.5051361,PARTHASARATHY2007313}), potentially introducing implicit size-dependent effects. Nevertheless, dislocation-density based CP models can effectively capture the microstructural evolution observed in small-scale materials without explicitly incorporating size-dependent terms in the constitutive equations.
This finding holds promising implications for the design of miniaturized mechanical devices and innovative materials in the rapidly evolving field of nanotechnology.

Building on our analytical treatment, we conduct numerical simulations of pillar compression with varying initial dislocation density  (i.e. initial microstructure) using an Eulerian crystal plasticity approach that effectively captures very large deformations. Our motivation stems from the burgeoning field of nanotechnology, where the production of nano-scale structures has underscored the importance of understanding material behavior at submicron levels~\cite{ref27:kurunczi2023avalanches}.
The compression test considered remains a pertinent example for illustrating insights derived from dislocation stability analysis, including the saturation of defect concentrations, and the impact of pre-existing microstructural features, while also linking these insights with experimental findings.

The rest of the paper is organized as follows.  In Section \ref{sec:Dislocation}, we present the dislocation-density-based flow rules.  In Section \ref{sec:attractors}, we study the stability conditions of  these rules and how they can be related to the initial microstructure.  Finally, in Section \ref{sec:numerics}, we introduce the Eulerian  2D model of crystal plasticity and we perform compression simulations on micro-pillar and explain our findings in terms of the stability of dislocation-density-based flow rules. 
In Section \ref{sec:conclusions}, we present our conclusions. 
\section{Dislocation-density based crystal plasticity models} \label{Chapter:model}
\label{sec:Dislocation}
\textcolor{black}{Dislocation-density-based crystal plasticity models provide a framework for describing the evolution of material strength by explicitly accounting for the interactions and dynamics of dislocations. These models relate the flow stress to the dislocation density and its evolution, capturing key mechanisms such as dislocation storage, annihilation, and dynamic recovery. Among the widely used models, the Kocks and Mecking (KM) model establishes a fundamental relation between strain hardening and dislocation density evolution. Extensions such as the Kocks-Mecking-Teodosiu and Kocks-Mecking-Estrin models introduce additional mechanisms to refine the prediction of strain hardening and recovery processes. More recently, a single arm source (SAS) controlled plastic flow model further advances this framework by incorporating enhanced descriptions of dislocation interactions~\cite{CUI2014279}. These models play a crucial role in simulating the mechanical response of crystalline materials under various loading conditions.}

\textcolor{black}{
In dislocation-density-based crystal plasticity models, the mechanical response of crystalline materials is governed by the interplay between elastic and plastic deformations. To accurately describe large plastic deformations, the deformation gradient is often multiplicatively decomposed into an elastic and a plastic part, enabling the formulation of constitutive laws within a geometrically consistent framework. The evolution of plastic flow is then characterized by flow rules that define the relationship between the plastic velocity gradient and the material behavior. These flow rules can be either rate-independent (plastic) or rate-dependent (visco-plastic), capturing different strain-rate sensitivities. The dislocation density evolution is directly linked to these flow rules, as the accumulation and annihilation of dislocations dictate the material's strain hardening or softening response. By incorporating dislocation-density-based hardening and recovery mechanisms, models such as the Kocks-Mecking, Kocks-Mecking-Teodosiu, Kocks-Mecking-Estrin, and CLLZ frameworks provide a physics-based approach to predict the evolving mechanical properties of crystalline materials under various loading conditions.}

\textcolor{black}{
In the following, we present this framework in detail, outlining the governing equations, flow rules, and dislocation-density-based hardening and softening mechanisms that define the material response.}

\subsection{Multiplicative decomposition of the deformation gradient}

Consider a single crystal at time \( t=0 \), free of any surface tractions and body forces. We choose this configuration, denoted as \( \mathcal{K}_0 \), as the reference configuration of the crystal. Let \( \mathcal{K} = \mathcal{K}(t) \) be the current configuration at time \( t \). The incorporation of lattice features is achieved through a multiplicative decomposition of the total deformation gradient \( \Fb \) into elastic and plastic components:  
\begin{equation}
    \Fb = \Fb^{e} \Pb.
    \label{eq:1}
\end{equation}  
This decomposition implies a two-stage deformation process. First, \( \Pb \) transforms the initial reference state \( \mathcal{K}_0 \) into an intermediate state \( \tilde{\mathcal{K}} \), which is characterized by plastic deformation only, with no change in volume. The tensor \( \Pb \) is referred to as the (visco)plastic deformation with respect to the reference configuration of a material neighborhood of the material point \( X \) at time \( t \). Then, \( \Fb^{e} \) brings the body to the final configuration \( \mathcal{K} \) through elastic deformation and rigid lattice rotation. Specifically, \( \Fb^{e} \) can be expressed as  
\begin{equation}
    \Fb^{e} = \Rb \Ub^{e},
\end{equation}  
where \( \Rb \) denotes the rotation of the crystal lattice with respect to its isoclinic orientation.

Following  \cite{Zeng2015},  
$\Pb$ is assumed to leave the underlying lattice structure undeformed and unrotated, ensuring the uniqueness of the decomposition in \eqref{eq:1}. The unique feature of CP theory is its construction of the plastic component $\Pb$ by constraining dislocation kinematics. Plastic flow evolves along pre-selected slip directions via volume-preserving lattice invariant shears, leaving the crystal lattice undistorted and stress-free  
\cite{Zeng2015,Nguyen2021, Baggio2019-rs}.

We label crystal slip systems  by integers $s= 1,...,N$, with $N$ denoting  the number of slip systems. Each slip system $s$ is  specified by the unit vectors $(\bbs_s^0, \bm_s ^0)$, where $\bbs_s^0$ is in the slip direction and $\bm_s^0$ is normal to the slip plane in the perfect undeformed lattice.  Since the viscoplastic deformation does not produce distortion or rotation of the lattice, the mean lattice orientation is the same in the reference and intermediate configurations and is specified by  $(\bbs_s^0, \bm_s^0 ), s= 1...N$.

The (visco)plastic deformation is due to slip only, with the slip contribution to the (visco)plastic deformation given by (\cite{ric71}, \cite{TeodSid76})  
\begin{equation}\label{flowrulenoi}
	\dot{\Pb} \Pb^{-1} = \sum_{s=1}^N \dot{\gamma}^{s} \bbs_s^0 \otimes \bm_s^0,
\end{equation}
where \( \dot{\gamma}^{s} = \dot{\gamma}^{s}(t) \) is the viscoplastic shear rate on slip system \( s \).

\subsection{Plastic and  visco-plastic flow rules }
\label{flowruled}

In order to complete the model, we need to provide the constitutive equation for the slip rate $\dot{\gamma_s}$ as a function of $\tau_s$, the stress  component acting on the slip plane of normal 
$\bm_s$ in the slip direction $\bbs_s$.
In the current configuration, $\tau_s$ is expressed as
\begin{equation}
	\tau^s = \bs : \M_s, 
\end{equation}
where $\bs=\bs(t)$ is the Cauchy stress tensor acting in the current configuration $\K$ while $\M_s$ is 
\begin{equation}\label{MR}
	\M_s= \left(\bbs_s\otimes\bm_s\right)^{symm}. 
\end{equation}
and $\bbs_s = \Fb^e \bbs_s^0$ and $\bm_s = (\Fb^e)^{-T} \bm_s^0$ are the slipping and normal  in the Eulerian configuration. 
 Note that  $\{\tau^s\}_{s=\overline{1,N}}$ are not independent; they belong to a fifth dimensional space  of $\R^N$ corresponding to the dimension of the space  of deviatoric stresses. 

To determine the shear strain rates $\dot{\gamma}_{\alpha}$ relative to the local stress, a constitutive law is needed. Various proposals exist, ranging from phenomenological to more physically based approaches. One simple phenomenological approach assumes that $\dot{\gamma}_{\alpha}$ depends on the stress only through the resolved shear stress $\tau^s$.

In rigid-plastic formulations, it is assumed that the onset of plastic flow of a slip system $s$ is governed by Schmid law:  the slip system $s$ is active if and only if $\vert \tau^s \vert=\tau_c^s $, i.e. 
\begin{equation}\label{Schmid}
	\dot{\gamma_s}(\vert \tau^s \vert -\tau_c^s)=0, \quad \dot{\gamma_s} \tau^s \geq 0, \quad \vert \tau^s \vert -\tau_c^s \leq 0, 
\end{equation}
where  $\tau_c^s$ is the slip resistance (also called critical resolved shear stress or CRSS).  For a given time $t$, the $\tau_c^s$ are material constants. Thus, the planes $\vert \tau^s \vert=\tau_c^s $ are the facets of the current yield surface of the single crystal in the stress space.   The associated internal plastic dissipation  functional is neither strongly convex  or  differentiable  and the solution could not be  unique. Additional assumptions are needed in order to restrict the number of solutions.

One way to overcome this difficulty of determining the active slip systems problem is to  adopt a rate-dependent approach for the constitutive response of the single crystal.
A widely used  rate-dependent (viscoplastic) model is the Norton-type model, which  relates the shear strain rate $\dot{\gamma}^{s}$ on a slip system $s$  to the resolved shear stress   $\tau^s $  through a power-law   (see Asaro and Needleman \cite{asaneed85}) 
\begin{equation}\label{shear_strain1}
 	\dot{\gamma}^{s}=\dot{\gamma}_{0}^{s}\,{\left\vert\frac{\tau^s}{\tau_c^s}\right\vert}^n\,\mbox{sign}(\tau^s),
\end{equation}
where $\dot{\gamma}_{0}$ is a reference shear strain rate,  while the exponent $\emph{n}$ has a fixed value.

Another regularization  of the Schmid law  can be done by using a  Perzyna-like viscoplastic law  of the form: 
\begin{equation}\label{flow_rule}
	\dot{\gamma}^s = \dfrac{1}{\eta_s}\left[\vert \tau^s \vert-\tau_c^s \right]_+ \mbox{sign}(\tau^s), 
\end{equation}
where $\eta_s$ is the  viscosity, which may depend on the slip rate,  and $[ \;  x\; ]_+=(x+|x|)/2\;$ denotes the positive part of any real number $x$. 
Note that the viscoplastic flow rule (\ref{flow_rule}) is the visco-plastic extension of the rigid-plastic Schmid law using an overstress approach.  The physical motivation for the dependence of the viscoplastic shear rate on the overstress $(\tau^s-\tau_c^s)$ was provided by Teodosiu and Sidoroff \cite{TeodSid76} based on an analysis of the microdynamics of crystals defects.

\subsection{Dislocation density hardening/softening models}  	

The yield limits $\tau_c^s$ of each slipping system $s$ can be considered as constants, but they can vary in time if hardening/softening effects are taken into consideration. A widely used  model  uses dislocation densities  on slip planes, denoted by $\rho^s$, as internal variables to link the microstructure to  macroscopic deformation.  To be more precisely, the critical resolved shear stress vector  $\btau_c=(\tau_c^{1}, ..,\tau_c^{N})$ depends on the dislocation densities vector $\bro=(\rho^1,..,\rho^N)$ as follows:
\begin{equation} \label{Hard}
	\btau_c =\bTau_c(\bro).  
\end{equation} 
For example, in the context of the Taylor model \cite{taylor1934mechanism}, and as proposed by Teodosiu and Raphanel \cite{TeodRafTab} (see also \cite{FRANCIOSI1980273}), the shear yield strength is expressed as :

\begin{equation} \label{HardTeo}
	\Tau_c^s(\bro)=\tau_0^s+ \alpha \mu b \sqrt{\sum_{p=1}^{N} d^{sp} \rho^{p}}, 
\end{equation} 
where $\tau_0^s$ is the friction stress, $\alpha$  is a 
dimensionless parameter  describing the mean strength of obstacles encountered by the mobile dislocation lines, $b$ is the modulus of the burgers vector parameter that relates the discrete plastic deformation of crystalline materials to dislocation motion, $\mu$ is is the shear modulus  and $d^{sp}$  is a dimensionless interaction matrix expressing the average strength of the interactions and reactions between slip systems.

\subsection{Dislocation density evolution models}  	

These internal variables represent microstructural features, with dislocation density being the most crucial in the context of plasticity. Dislocations are the primary carriers of plastic deformation, making dislocation density a key microstructural state variable.

The evolution of the  dislocation density  can be developed from a  phenomenological (statistical) model. The dislocation density \textcolor{black}{rate} 	$\dot{\rho}^{s}$, which is proportional to the plastic slip rate $\vert \dot{\gamma}^{s}  \vert$, is  the sum of  the  dislocation multiplication rate $\MM^s$ and of  the dislocation annihilation rate $-\Ac^s$: 
\begin{equation} \label{rhoEv}
	\dot{\rho}^{s} = \frac{1}{b} \left(\MM^s(\bro)-\Ac^s(\bro)\right) \vert \dot{\gamma}^{s}  \vert. 
\end{equation}
For instance in the model of Kocks-Mecking-Teodosiu  \cite{MECKING1970427, TeodRafTab}  the multiplication and  annihilation rates  are given by   
\begin{equation} \label{KM}
	\MM^s(\bro)=\frac{\sqrt{\sum_{p=1}^{N} a^{sp} \rho^{p} }}{k}, \quad \Ac^s(\bro)=2y_c \rho^{s}, 
\end{equation}
where 	$y_c$ is the mean distance controls the annihilation of dislocations and $k$ is a proportionality factor, it represents the number of obstacle before a mobile dislocation get stopped and  $a^{sp}$  is a dimensionless interaction matrix. 
The values of interaction $a^{sp}$ coefficients are listed in the Table \ref{tab:FCC_asp_Ni}, which were determined from dislocation dynamics simulations \cite{Devincre2008-to, kubin2008modeling}.
\begin{table}[h]
	\begin{center}
		\begin{tabular}{c c c c c c} 
			\hline \hline\\
			$a_{0}^{self}$& $a_{cop}^{coplanar}$ & $a_{orth}^{Hirth}$& $a_{gli}^{glissile}$ & $a_{lom}^{lomer}$ & $a_{col}^{collinear}$ \ \\
			\hline \hline\\
			0.122 & 0.122 &0.07&0.137 &0.127 &0.625 \\
			\hline \hline 
		\end{tabular}
	\end{center}
	\caption{Values of the six independent interaction coefficients $a_{ij}$ for FCC crystals.}\label{tab:FCC_asp_Ni}  
\end{table}

In the case of  generalized  Kocks-Mecking-Estrin model \cite{YASNIKOV2022142330}, which is \textit{size-dependent}, the expressions for dislocation multiplication $\MM^s$ and annihilation rates $\Ac^s(\bro)$ are provided as follows : 
\begin{equation} \label{KME}
\MM^s(\bro)=\frac{\tilde{K}_0}{D}+\tilde{K}_1 \sqrt{\rho^{s}}, \quad \Ac^s(\bro)= b K_2 \rho^{s}, 
\end{equation} 
where    $\tilde{K}_0$ is a production rate controlling factor (dimensionless),  $D$ the average grain size,  $\tilde{K}_1$ governs the rate of dislocation storage by the interaction with immobile dislocations and ${K}_2$ is a controlling factor of recovery by the annihilation of dislocations.

In an alternative size-dependent model called the CLLZ model \cite{CUI2014279}, the multiplication and  annihilation rates  are :
$$\MM^s(\bro)=\frac{1}{2\bar{\lambda}}+k_{f}\sqrt{\rho^{s}}, \quad \Ac^s(\bro)= y \rho^{s} + \frac{2 \cos^2(\beta/2)}{D},$$  
where the model parameters  are :  $\bar{\lambda}$ is the length of the statistically average effective Single Arm Source, $k_{f}$ is a dimensionless constant and set at $10^{-2}$,  $y$ is the effective mutual annihilation distance and set to be equal to $6b$, $\beta$ the slip-plane orientation angle is the angle between the primary slip plane and the top surface of the single-crystal micro-pillars and  $D$ is the diameter sample.

In the dislocation density-based constitutive model of  Kameda and Zickry \cite{KAMEDA1998631} it 
is assumed that  the  total dislocation-density  $\rho^s$  can be  decomposed, into a mobile dislocation-density, $\rho_{m}^{s}$, and an immobile dislocation-density  $\rho_{im}^{s}$ as
$\rho_{tot}^{s}=\rho_{m}^{s}+\rho_{im}^{s}.$ 
For both of them  we deal with two evolution equations 	
$$\dfrac{d \rho_{m}^{s}}{dt}=\left(\dfrac{{g_{sour}}}{b^{2}} \dfrac{ \rho_{im}^{s}}{\rho_{m}^{s}}-\dfrac{{g_{min ter}}}{b^{2}} \exp(-\dfrac{{H}}{kT})-\dfrac{{g_{immob}}}{b}\sqrt{\rho_{im}}^{s}\right)\vert \dot{\gamma}^{s}\vert,$$
$$
\dfrac{d\rho_{im}^{s}}{dt}=\left(\dfrac{{g_{min ter}}}{b^{2}} \exp(-\dfrac{{H}}{kT})+\dfrac{{g_{immob}}}{b}\sqrt{\rho_{im}^{s}}-g_{recov} \exp(-\dfrac{{H}}{kT}) \rho_{im}^{s}\right)\vert \dot{\gamma}^{s}\vert$$
where $ g_{sour}$ represents the coefficient associated with the rise in mobile dislocation density attributed to dislocation sources, $ g_{min ter}$ denotes the coefficient linked to the trapping of mobile dislocations resulting from interactions with forest intersections, cross-slip mechanisms around obstacles, or dislocation interactions, $ g_{recov}$ signifies the coefficient related to the rearrangement and annihilation of immobile dislocations, $ g_{immob}$ denotes the immobilization of mobile dislocations, $H$ is the activation enthalpy, and $k$ is Boltzmann's constant.  Since only the  immobile dislocation densities  are  responsible of the hardening/softening effect and their evolution is independent of the mobile ones we can conclude that  immobile dislocation-density  $\rho_{im}^{s}$ plays the same role as  $\rho^s$ in the previous models. Moreover,    the equation of immobile dislocation density evolution  has the same structure as the generalized  Kocks-Mecking-Estrin model. For that we have to take $\displaystyle \frac{\tilde{K}_0}{D_m}=\dfrac{{g_{min ter}}}{b} \exp(-\dfrac{{H}}{kT}), \;  \tilde{K}_1 =  g_{immob}$ and $K_2= g_{recov} \exp(-\dfrac{{H}}{kT}) $.  

\bigskip 

Finally,  the hardening or softening  effects,  described by (\ref{Hard}) and (\ref{rhoEv}),  depend on  the  evolution  of the dislocation density $\rho^{s}$ on all systems $s$, that derives from the balance between accumulation and annihilation rates.

\section{Dislocation density attractors in slip  driven processes} \label{Chapter:DD}
\label{sec:attractors}

We  aim here  to analyze the stability of several established dislocation density-based models, including the Kocks and Mecking (KM) model, the Kocks-Mecking-Teodosiu model, the Kocks-Mecking-Estrin model, and the CLLZ model. The objective is to identify the conditions under which saturation states emerge in active slip systems. Furthermore, we assess the linear stability of these states to determine whether they exhibit attractor-like behavior.

\subsection{Stability analysis and attractors}

In this section we will consider a slip   driven process, which  in our context means that the slip functions $t\to \gamma^s(t)$ are given for a significant time period $[0,T]$ for all $s=1,..,N$.  The $N$ slip systems   can be classified  between  $N_a$ active slip systems  and $N_i$ inactive slip systems ($N=N_a+N_i$). To be more precisely  let $A_c \cup A_i$ be a partition of $\{1,..,N\}$ and let us denote by $\bro_a$ the vector containing all  active slip systems $\rho^s, s\in A_c$, for which $\vert \dot{\gamma}^{s}  \vert >0$, and by  $\bro_i$  the vector with all   inactive slip systems $\rho^s, s\in A_i$,  for which $\dot{\gamma}^{s} =0$. Following (\ref{rhoEv}) the dislocation densities of inactive slip systems will rest constant, i.e. $\bro_i(t)=\bro_i(0)=\bro_{i}^0$, while the active ones will satisfy the following Cauchy problem 
\begin{align}
	\left\{\begin{array}{cl}\label{rhoEvAct}
		\dot{\rho}^{s}(t) =& \displaystyle \frac{1}{b} \left(\MM^s(\bro_a(t),\bro_{i}^0)-\Ac^s(\bro_a(t),\bro_{i}^0)\right) \vert \dot{\gamma}^{s}(t)  \vert,    \\ 	\rho^s(0)=&\rho_0^s, \quad s\in A_c.
	\end{array}\right.
\end{align} 
Let $\dot{\gamma}^{ref}$ be a reference slip rate and let  denote by $\gamma=\dot{\gamma}^{ref}t$ the reference slip.  For simplicity we consider only constant slip rates, i.e. $\dot{\gamma}^{s}(t) =
\dot{\gamma}^{ref}g_s$. Since the  model considered here are time independent the differential system  (\ref{rhoEvAct}) can be recasted in terms of $\gamma$ as 	
\begin{align}
	\left\{\begin{array}{cl}\label{rhoEvActGam}
		\displaystyle	\frac{d\rho^s} {d \gamma}=&\displaystyle  \frac{1}{b} \left(\MM^s(\bro_a(\gamma),\bro_{i}^0)-\Ac^s(\bro_a(\gamma),\bro_{i}^0)\right) \vert g^s\vert,    \\ 	\rho^s(0)=&\rho_0^s, \quad s\in A_c.
	\end{array}\right.
\end{align} 
\textcolor{black}{To describe the attractors of the dislocation density (also referred to as the "saturation dislocation density"), we first need to compute the stationary (or invariant) dislocation densities, denoted by the vector $\tilde{\bro}_a$, as the solution of the following system:}
\begin{equation} \label{Sat}
	\MM^s(\tilde{\bro}_a,\bro_{i}^0)=\Ac^s(\tilde{\bro}_a,\bro_{i}^0),  \quad s\in A_c.
\end{equation}
We remark that if  $\bro_a^0=\tilde{\bro}_a$ then $\bro_a(t)=\tilde{\bro}_a$ for all $t\in [0,T]$. The attractors are linearly stable  stationary  dislocation densities. 
One can characterize the stability of  $\tilde{\bro}_a$ through the eigenvalues $\lambda^q(\tilde{\bro}_a,\bro_{i}^0), q=1,..N_a$ of the matrix 
$$ S_{sp}=(\frac{\partial\MM^s}{\partial\rho^p}(\tilde{\bro}_a,\bro_{i}^0) -\frac{\partial\Ac^s}{\partial\rho^p}(\tilde{\bro}_a,\bro_{i}^0)) \vert g^{s} \vert,\quad s,p\in A_c.$$     
If 
\begin{equation} \label{Lamb}	
	Re(\lambda^q(\tilde{\bro}_a,\bro_{i}^0))<0 \quad \mbox{for all} \; q=1,..N_a  
\end{equation} 
then $\tilde{\bro}_a$ is linearly stable and it is an attractor, called in the next {\em saturation  dislocation density} and denoted by $\bro_a^{sat}$. That means that there exists a  neighborhood ${\cal N}$  of  $\bro_a^{sat}$ such that if $\bro_a^0 \in {\cal N}$ then $\bro_a(t) \to  \bro_a^{sat}$ and $\btau_c(t)\to \btau_c^{sat}= \bTau_c(\bro_a^{sat},\bro_{i}^0)$.

\label{sec:sizeeffects}
\subsubsection{Self interaction dislocations} 

To continue, we first consider the case where the dislocation interactions are limited to self interaction, which means that the dislocation multiplication and annihilation rates for the slip system $s$ depend only on  $\rho^s$, i.e. $\Tau^s(\bro)=\Tau^s(\rho^s), \MM^s(\bro)=\MM^s(\rho^s), \Ac^s(\bro)=\Ac^s(\rho^s)$ for all systems $s$.   For  Kocks and Mecking's model  (\ref{KM})  "self interaction" means  $d^{sp}=a^{sp}=0$ for $s\neq p$. 

For self interaction dislocations is no longer necessary to distinguish between the active and inactive systems as in the previous section.  To find the stationary dislocation densities $\tilde{\rho}^s$ the nonlinear  \textcolor{black}{system} (\ref{Sat})   reduces to a nonlinear equation 
\begin{equation}\label{SatSI}
	\MM^s(\tilde{\rho}^s)=\Ac^s(\tilde{\rho}^s)
\end{equation}	
while the condition (\ref{Lamb}), which assure  that a stationary density is an attractor (or a saturation density),  reads 
\begin{equation}\label{LambSI}
	\frac{d}{d\rho}\MM^s(\tilde{\rho}^s)<\frac{d}{d\rho}\Ac^s(\tilde{\rho}^s).   \end{equation}

\bigskip

In the next we analyze  three specific models: (KM),  (KME) and (CLLZ).

\bigskip

i) For  {\em Kocks and Mecking's model}  (\ref{KM}) we found from (\ref{SatSI}) two stationary density values for each slip system 
$\tilde{\rho}^s=0, \quad \mbox{and} \quad  \tilde{\rho}^s= a^{ss}/(2y_{c}k)^2=a_{0}^{self}/(2y_{c}k)^2,$  
but according to (\ref{LambSI}) only the second one is stable that means that we deal with a single  attractor or saturation dislocation density  $\rho_{sat}^{s}$ which corresponds to a  saturation CRSS $\tau_{c,sat}^{s}$.  
\begin{equation}\label{KM-self}
	\rho_{sat}^{s} =\frac{ a^{ss}} {({2y_{c}k})^2}, \quad  
    \tau_{c,sat}^s = \tau_0^s + \frac{\alpha \mu b \sqrt{d^{ss} a^{ss}}}{2 y_c k}
\end{equation}
\noindent

\begin{table}[h!]
	\centering
	\renewcommand{\arraystretch}{1.2}
	\begin{tabular}{c c c c c c c}
		\hline \hline
		$\tau_0$ & $\alpha$ & $\mu$ & $b$ & $y_c$ & $k$   \\
		\hline \hline
		11 MPa& 0.25 & 76 GPa & 0.245 nm & $3.36\times b$ & 38  \\
		\hline \hline
	\end{tabular}
	\caption{The parameters of DD hardening/softening model for a FCC single crystal.}\label{tab:FCC_Ni}
\end{table}

\begin{figure}[htbp]
	\centering
	\begin{minipage}[b]{0.48\textwidth}
		\centering
		\includegraphics[scale=0.3]{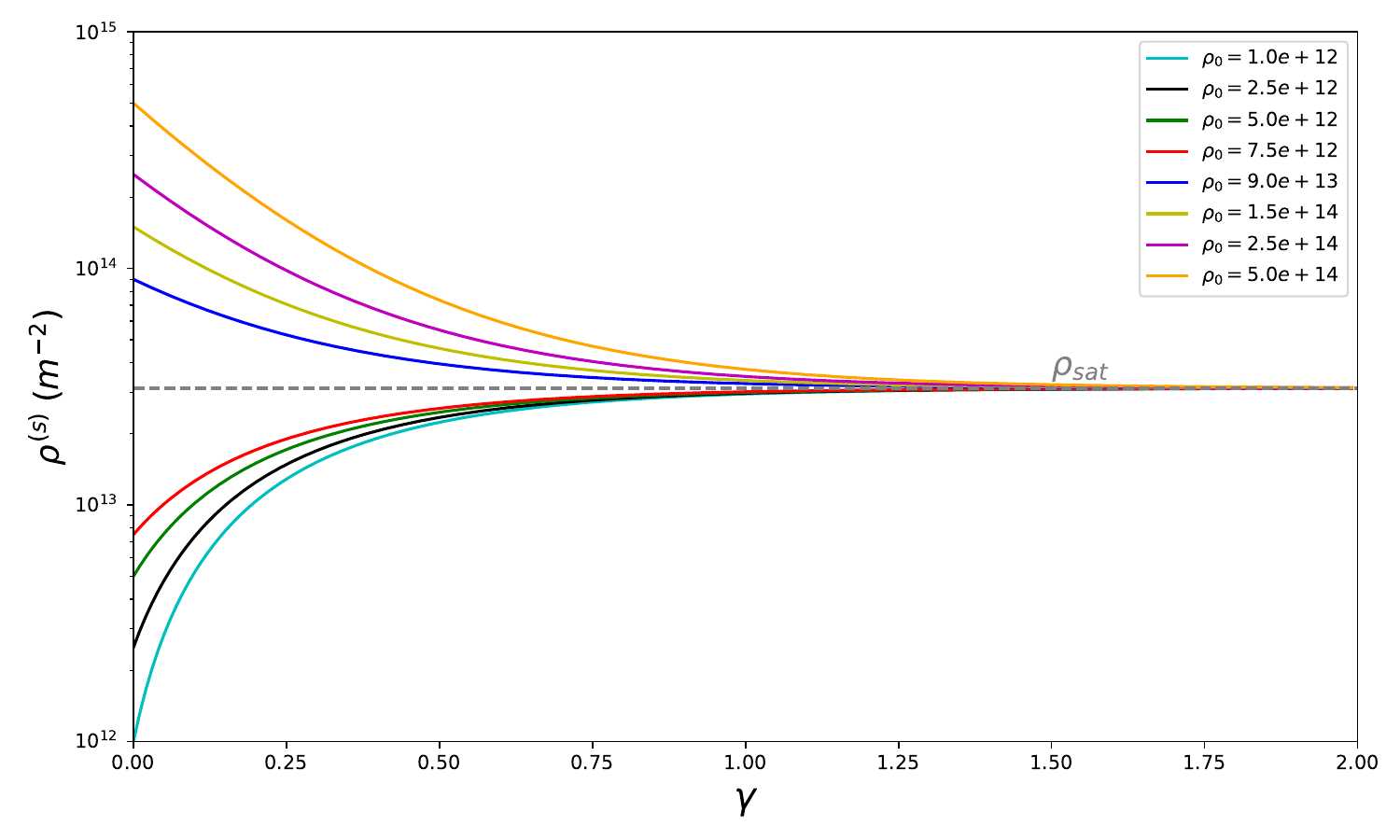}
	\end{minipage}
	\hfill
	\begin{minipage}[b]{0.48\textwidth}
		\centering
		\includegraphics[scale=0.3]{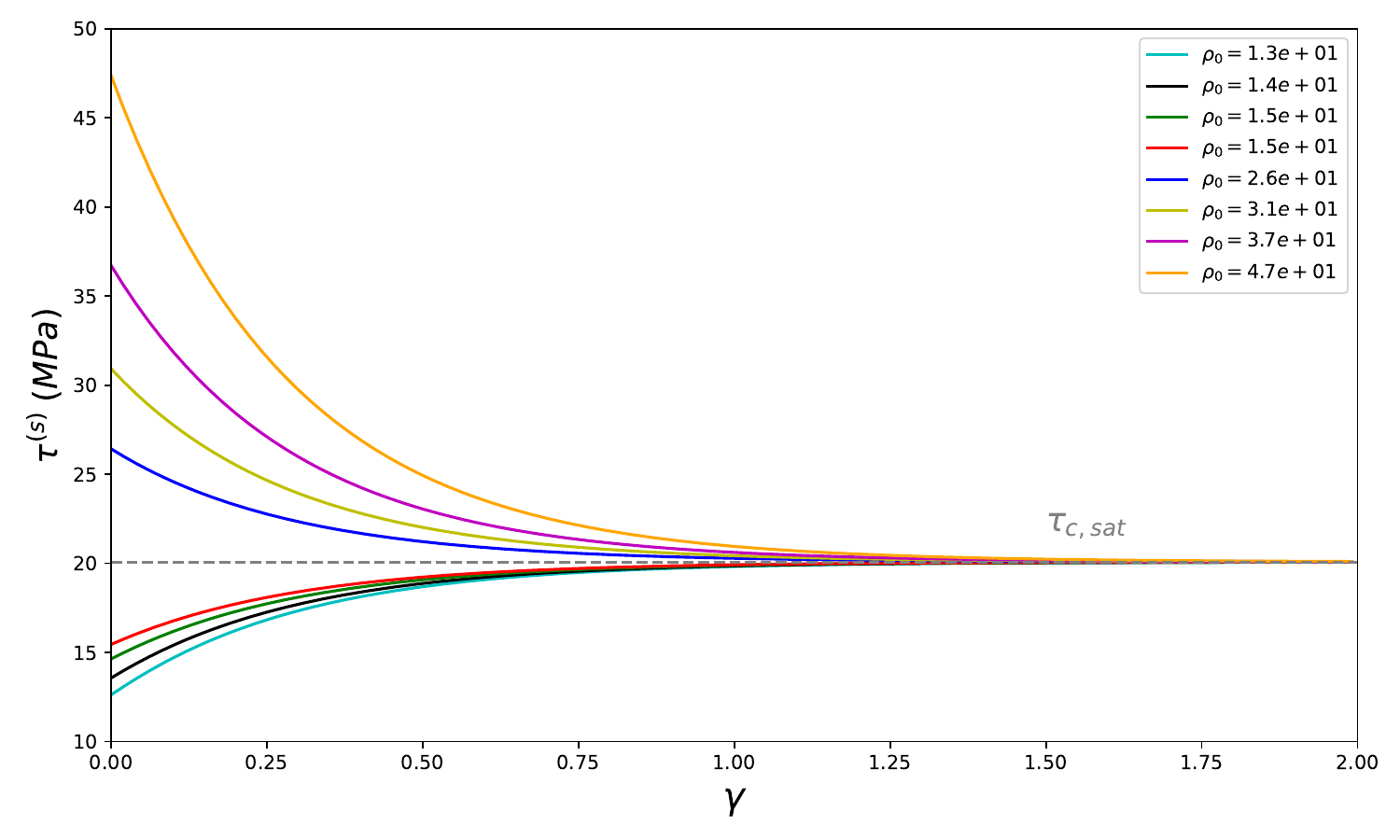}
	\end{minipage}
	
	\caption{Evolution of a  self-interacting system  for eight different initial values $\rho^s_0$ computed  for the  KM model. Left: dislocation density  (in m$^{-2}$)  over slip $\gamma  \to \rho^s(\gamma)$. Right: shear yield strength (in MPa) over slip $\gamma  \to \tau^s_c(\gamma)$.}
	\label{Self}
\end{figure}

As an example, we have computed in Fig.~\ref{Self} (left) the evolution of the dislocation density $\gamma \to \rho^s(\gamma)$ (i.e., the solution of the Cauchy problem~(\ref{rhoEvActGam})) for a FCC single crystal. Physical values, summarized in Table~\ref{tab:FCC_Ni}, are taken from \cite{PARTHASARATHY2007313,ALCALA20083277}. We consider 8 choices of the initial dislocation densities $\rho^s_0 \in [1 \times 10^{12},\,5 \times 10^{15}]$, and set $a^{s,p} = 0$ for $s \neq p$, leading to the saturation values $\rho_{sat}^{s} = 3.1169 \times 10^{13}$ and $\tau_{c,sat}^{s} = 20.0774~\text{MPa}$, as mentioned in equation~(\ref{KM-self}).  Since  the saturation value of dislocation density and CRSS is uniform across the gliding systems the plot concerns only  one slip system with different initial DD.

	
Given the similarity between interaction matrix $a^{s,p}$ and $d^{s,p}$, we have assumed  that  $a^{s,p}=d^{s,p}$. 
We remark that the dislocation densities and the critical resolved shear stress are converging to saturation values, as predicted by the theory, with a softening effect for   $\rho_{0}^{s}>\rho_{sat}^{s}$ and a hardening one for  $\rho_{0}^{s}<\rho_{sat}^{s}$. Moreover, this convergence is effective for large values of slips (more than 50$\%$), which are expected  in the shear bands for instance.

\bigskip 

ii) For  {\em generalized  Kocks-Mecking-Estrin (KME) model}  \cite{estrin1984unified} the formula of the saturation dislocation density is
\begin{equation}\label{KMEself}
	\rho_{sat}^{s} = \left(\frac{\tilde{K}_{1} + \sqrt{\tilde{K}_{1}^{2}+\; 4 \; b \; K_{2} k_{0} }}{2 \; b \; K_{2}}\right)^{2}
	\quad \text{with} \quad  k_{0} = \frac{\tilde{K}_0}{D} 
\end{equation}

\bigskip 

iii) For the{em CLLZ model}~\cite{CUI2014279}, the saturation dislocation density and corresponding critical resolved shear stress (CRSS) are given by:
\begin{equation}\label{CLLZ}
	\rho_{sat}^{s} = \left( \frac{ k_{f} + \sqrt{k_{f}^{2} +\; \frac{4 \; b \; y}{D} \left[ \frac{1}{2\xi} - 2 \cos^2(\beta/2) \right] } }{2 y} \right)^{2},
\end{equation}	 
\begin{equation}
	\tau_{c,sat}^{s} = \tau_0^s + \alpha\mu b \sqrt{ \rho_{sat}^{s} } + \frac{k \mu b}{\bar{\lambda}} 
	\quad \text{with} \quad \xi = \frac{\bar{\lambda}}{D} \approx 0.3.
\end{equation}

\begin{figure}[h]
	\makebox[\textwidth][c]{\includegraphics[scale=0.4]{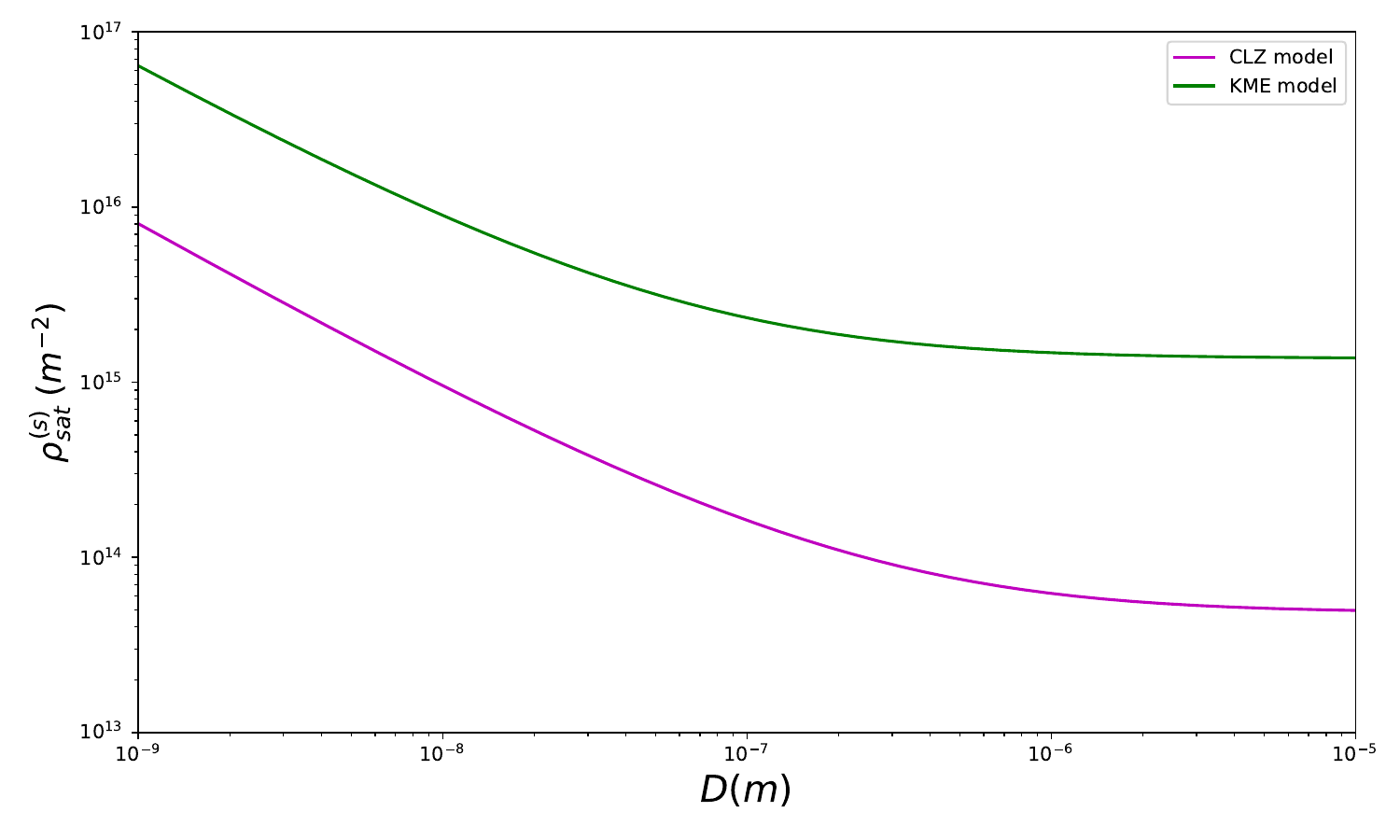}}
	\caption{Saturation dislocation density $\rho_{sat}^{s}$ (in m$^{-2}$) predicted by the KME model (\ref{KMEself}) and the CLLZ model (\ref{CLLZ}) as a function of the scale parameter $D$(in m).}\label{rhosat_KME_CLZ}
\end{figure}

\bigskip

Figure~\ref{rhosat_KME_CLZ} illustrates the influence of grain size $D$ on the saturation dislocation density $\rho_{sat}^{s}$ for the following settings: $b = 0.24$ nm, $\tilde{K}_1 = b \times 1.07 \times 10^{10}$, $k_2 = 289.63$, $\tilde{K}_0 = 3.8$ for (KME) model and $b = 0.24$ nm, $y = 6b$, $k_f = 0.01$, $\beta = 49^\circ$ for (CLZZ) model. A pronounced size-effect is observed for both the KME and CLLZ models when the specimen diameter is below $10^{-7}~\text{m}$, where $\rho_{sat}^{s}$ decreases rapidly as the diameter increases. Beyond this threshold, the size-effect becomes minor: the variation in $\rho_{sat}^{s}$ slows down considerably, and both models tend toward a plateau near $5 \times 10^{-6}~\text{m}$, where further increases in diameter have negligible impact. The angle $\beta$ used in the CLLZ model corresponds to values measured in single-slip micro-pillar compression experiments. In the range $D \in [200~\text{nm}, 10~\mu\text{m}]$, relevant to such micro-pillar tests, the saturation dislocation density predicted by the three models (KM, KME, and CLLZ) remains largely insensitive to the grain size.

 \bigskip

In conclusion for self interaction dislocations we can expect three  different scenarios, depending of the initial condition  $\rho_{0}^{s}$ of the density dislocation: (i) If 	$\rho_{0}^{s}<\rho_{sat}^{s}$ then  $\rho^s(t) \nearrow \rho_{sat}^{s}$ and we deal a with a hardening process  $\tau_c^s(t) \nearrow \tau_{c,sat}^{s}$ up to a maximal critical resolved shear stress;  (ii)  if  $\rho_{0}^{s}>\rho_{sat}^{s}$ then  $\rho^s(t) \searrow \rho_{sat}^{s}$ and we deal a with a softening  process  $\tau_c^s(t) \searrow \tau_{c,sat}^{s}$ down to a minimal critical resolved shear stress; (iii) if  $\rho_{0}^{s}\approx\rho_{sat}^{s}$ then  $\rho^s(t) \approx \rho_{sat}^{s}$ and we deal a with a plateau   $\tau_c^s(t) \approx \tau_{c,sat}^{s}$.

\subsubsection{Cross-interacting dislocations} 
\textcolor{black}{We now examine the Kocks and Mecking model as it applies to cross-interacting dislocations. This framework extends beyond self-interactions to account for the complex interplay between dislocations on different slip systems.}  The nonlinear system   (\ref{Sat})  for stationary dislocation densities $\tilde{\bro}_a$ reads:  
\begin{equation} \label{SatKM}
	\sum_{p\in A_c} a^{sp}\tilde{\rho}^{p} +\rho_{0,i}^s=(2y_ck)^2( \tilde{\rho}^{s})^2,    \quad s\in A_c,
\end{equation}
where 
$$\rho_{0,i}^s= \sum_{p\in A_i} a^{sp}\rho^{p}_0,  \quad s\in A_c.$$
Generally,  the above system cannot  be solved analytically and a numerical approach is needed. For instance one can use a Newton-Raphson method or to try to compute the local minimizers of   the potential function 
$$W(\bro_a)=\frac{(2y_ck)^2}{3}\sum_{p\in A_c} (\rho^{p})^3-\frac{1}{2} \sum_{s,p\in A_c} a^{sp}\rho^{p}\rho^{s}-\sum_{s\in A_c,p\in A_i} a^{sp}\rho^{p}_0\rho^{s}.$$ 
For the simplicity of the analysis we will suppose in the next that  the slip rates $\dot{\gamma}^{s}=\dot{\gamma}^{ref}$ are the same in all active systems  $s\in A_c$, i.e. $g^s=1$. To see if a solution $\tilde{\bro}_a$ is an attractor we have to compute the eigenvalues of  $\tilde{\alpha_s}$ of the  matrix
$$ \tilde{A}_{sp}=\frac{a^{sp}}{\tilde{\rho}^{s}}.$$ 
Then the linear stability condition (\ref{Lamb})  reads 
\begin{equation} \label{LambKM}
	Re(\tilde{\alpha_q})<
	2(2y_ck)^2,   \quad  \mbox{for all} \; q=1,..,N_a.
\end{equation}
Let us mention here a sufficient  condition which allow us to compute an analytic  solution of (\ref{SatKM}). Let  suppose  that the sum of all cross-interaction coefficients  corresponding to the active slip systems $ a^{sp}$ on a line $s$  is the same for all lines, i.e. 
\begin{equation} \label{Sa}
	\sum_{p\in A_c} a^{sp} =a_c, \quad \rho_{0,i}^s=\rho_{0,i},  \quad  \mbox{for all} \; s\in A_c. 
\end{equation} 

 This assumption is satisfied for FCC crystals having their 12 slip systems active  or specific active slip systems configurations. Supporting data examples are given in \cite{dequiedt2015heterogeneous,madec2017dislocation}.

If the above property is verified then there exists a uniform   solution of  (\ref{SatKM}), given by $\tilde{\rho}^{s}=\tilde{\rho}$ for all $s\in A_c $, where $\tilde{\rho}$  is 
$
	\tilde{\rho}=(a_c\pm \sqrt{a_c^2+4{({2y_{c}k})^2\rho_{0,i}}})/(2(2y_{c}k)^2). 
$
If  $\rho_{0,i}$ is not vanishing then only one solution (corresponding to the sign $+$) is non-negative. 
If  $\rho_{0,i}=0$ (this is always the case if all slip systems are active $A_i=\emptyset$ or if  the active and inactive systems are not interacting, i.e.    $a^{sp}=0$ for all $s\in A_c, p\in A_i$) then we deal with two uniform stationary dislocation densities:  
$	\tilde{\rho}=0 \quad \mbox{and} \quad  \tilde{\rho}=a_c/({2y_{c}k})^2$.

To see if the stationary uniform density is an attractor we have to check the stability condition (\ref{LambKM}).  We remark that if (\ref{Sa}) holds then  the largest eigenvalue of  matrix  $ \tilde{A}_{sp}$, corresponding to the uniform eigenvector $(1,..,1)$, is 
$\tilde{\alpha}^{max}=a_c/\tilde{\rho}$ and  (\ref{LambKM}) holds always  for $\tilde{\rho}>0$.  We deduce that for Kocks-Mecking model we deal with  a  uniform saturation dislocation density but   (generally) a  non-uniform saturation critical resolved shear stress,   given by 
\begin{equation} \label{Rosatcross}
	\rho_{sat}=\frac{a_c + \sqrt{a_c^2+4{({2y_{c}k})^2\rho_{0,i}}}}{2(2y_{c}k)^2},\quad 	\tau_{c,sat}^{s}=\tau_0^s+\alpha \mu b \sqrt{{\color{sepia}{\rho}_{sat}} \sum_{p\in A_c} d^{sp} + \sum_{p\in A_i} d^{sp} \rho^{p}_0}.
\end{equation}

\begin{figure}[!hbt]
    \centering
    \begin{minipage}[b]{0.48\textwidth}
        \centering
        \includegraphics[scale=0.3]{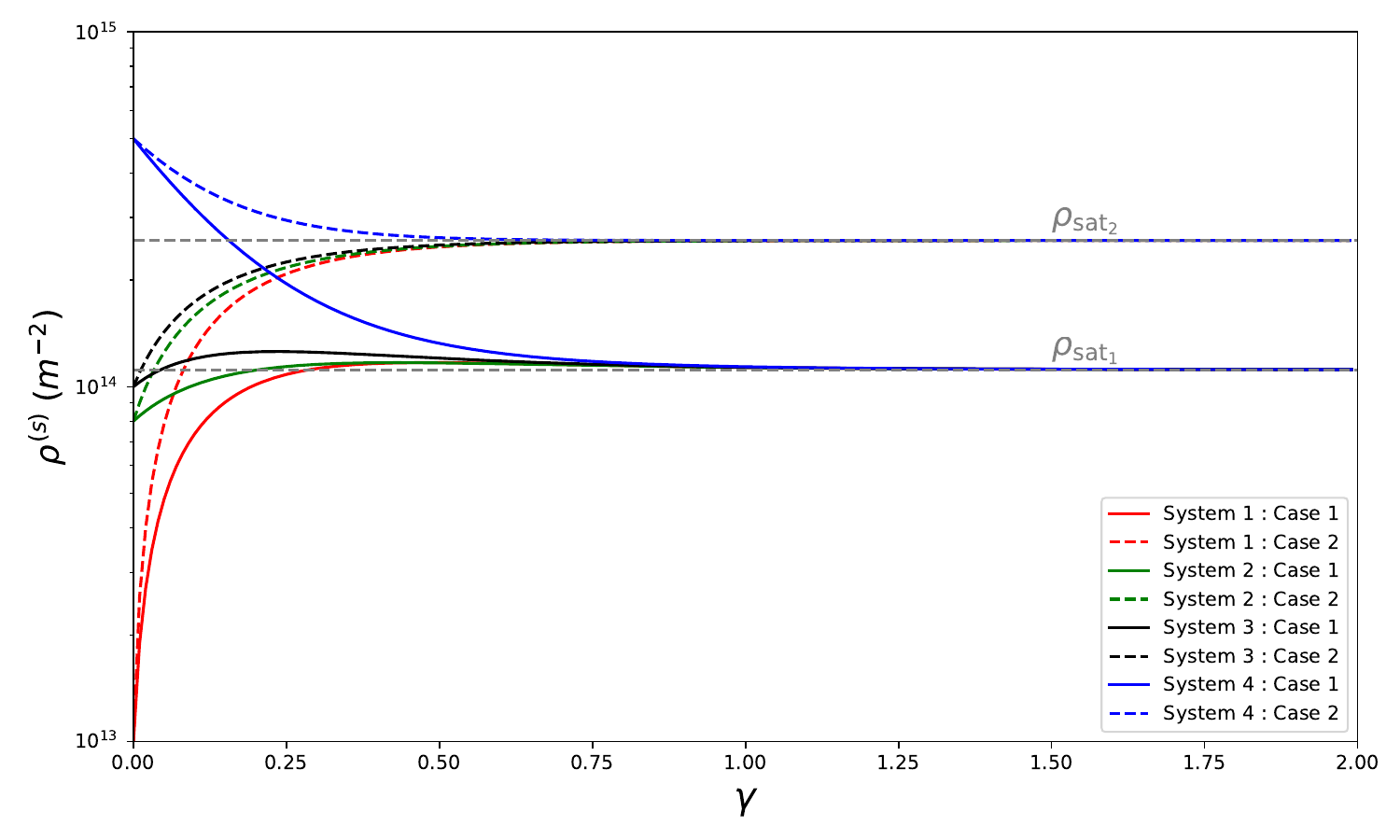}
    \end{minipage}
    \hfill
    \begin{minipage}[b]{0.48\textwidth}
        \centering
        \includegraphics[scale=0.3]{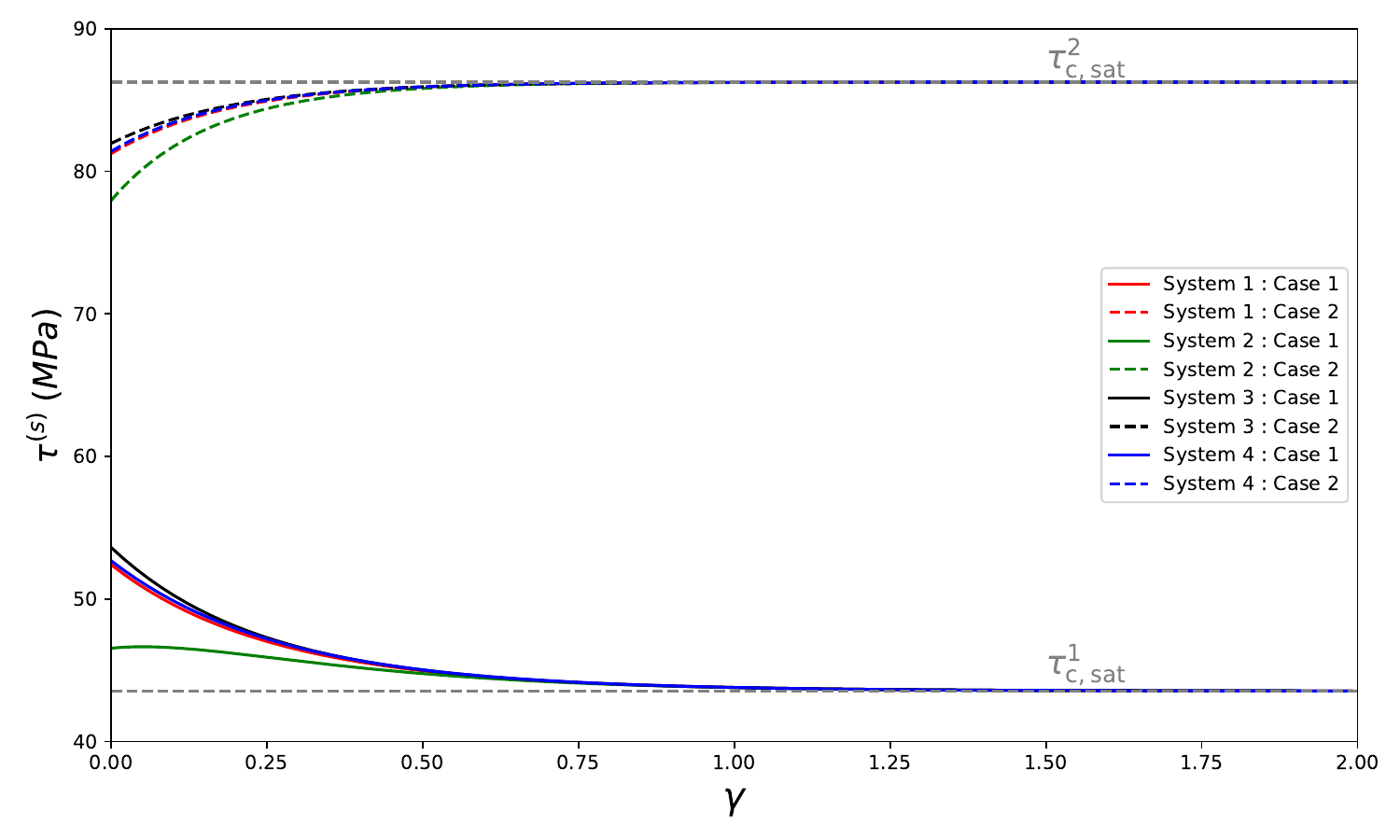}
    \end{minipage}

    \caption{Evolution of a FCC single crystal cross interacting system (KM) model  (\ref{Rosatcross})   for four different initial values $\rho^s_0$ and  two choices ($\alpha$) and ($\beta$) of the level of inactive dislocation densities. Left: dislocation density  (in m$^{-2}$) over slip $\gamma  \to \rho^s(\gamma)$. Right: shear yield strength (in MPa) over slip $\gamma  \to \tau^s_c(\gamma)$, for $s\in A_c$.}
    \label{Cross}
\end{figure}

In order to see the role played by the level of inactive dislocation densities, denoted here by \(\rho_{0,i}\), we have computed in Fig.~\ref{Cross} (left)  the evolution of the dislocation density \(\gamma^s \to \rho^s(\gamma^s)\) (i.e., the solution of the Cauchy problem (\ref{rhoEvActGam})) for a single FCC crystal  in the case of four active slip systems (denoted by $s=1,2,3,4$, i.e. $A_c=\{1,.,4\}, A_i=\{5,..,12\}$), considering two choices of inactive dislocation densities.

\begin{itemize}
	\item ($\alpha$) $\rho^{p}_0=10^{11} m^{-2}$ for all $p\in A_i$
	\item 	 ($\beta$) $\rho^{p}_0=10^{14}  m^{-2}$, for all $p\in A_i$. 	
\end{itemize}

The interaction types between the four active slip systems are such that $(1,4)$ and  $(2,3)$ are Lomer, 
$(1,3)$ and $ (2,4)$ are orthogonal (Hirth) , and $(1,2)$ and  $(3,4)$ are coplanar.  We remark that the assumption (\ref{Sa}) is satisfied and the sum of all cross-interaction coefficients corresponding to the active slip systems is  $a_c = a_{\text{cop}}^{\text{coplanar}} + a_{\text{orth}}^{\text{Hirth}} + a_{\text{lom}}^{\text{Lomer}}$, with $a^{sp} \approx d^{sp}$, values are given in Table~\ref{tab:FCC_asp_Ni}. The other parameters ($y_c, k, \tau_0, \alpha, \mu$, and $b$) are listed in Table~\ref{tab:FCC_Ni}.

Thanks to the different initial dislocation density values, the initial values of \( \tau^s_c \) are not identical across the active slip systems \( s \in A_c \). Let us remark that, in our particular case, the four active slip systems evolve toward a uniform  critical resolved shear stress (CRSS). This  is due the fact that the sum of all cross-interaction coefficients \( a^{sp} \) along each active slip system \( s \) is identical for all four systems.

We have considered four uniform  initial values of the four active  dislocation densities  $\rho^s_0= 1 \times 10^{13}, \rho^s_0=8 \times 10^{13},  \rho^s_0=1 \times 10^{14}$ and $\rho^s_0= 5 \times 10^{14}]$m$^{-2}$ for all $s\in A_c$. 
We remark that the dislocation densities and the  critical resolved shear stress are converging to saturation values $\rho_{\text{sat}} =  1.1307 \times 10^{14}$m$^{-2}$, $\tau_{sat}=43.5395$MPa for ($\alpha$) and  $\rho_{sat}=2.584 \times 10^{14}$m$^{-2}$,  $\tau_{sat}=86.2565$ MPa for ($\beta$), as is predicted by the theory. The value of the saturation is strongly dependent on the values of inactive dislocation densities. The level of inactive dislocation densities is essential in the evolution of yield limit. Indeed, as shown in Fig. \ref{Cross} (right), in the first case we deal with a hardening process while in the second one we observe a softening behavior.

\section{Saturation dislocation density in pillars  compression} 
\label{sec:numerics}

To see how the stability analysis presented earlier can provide insight into practical mono-crystal plasticity deformation processes, we conduct in this section numerical simulations of pillar compression.   These simulations aim to demonstrate that the behavior of mono-crystal pillars under compression tests is strongly dependent on the initial dislocation density which could  depend on the pillar size, particularly noticeable in FIB-prepared samples where the initial dislocation density decreases as the sample diameter increases~ \cite{Jennings2010-yg}.

\subsection{The  Eulerian  crystal plasticity 2D model} \label{ToyModel}
\label{sec:euler}

The choice of  the pillar compression processes to illustrate the above theory is due to  the fact that   the saturation  dislocation density is effective only for large strains.   We briefly describe here  a simple 2D Eulerian crystal plasticity model  (see \cite{Cazacu2010-hi}),  that can capture very large deformations.  It is acknowledged that the reduction from a 3D model to a 2D model, which incorporates only three slip systems, may result in the loss of some detailed information. Despite this simplification, the derived conclusions can still offer significant insights into the underlying physical phenomena. 

\subsubsection{Rigid-(visco)-plastic model} 
Since in applications involving large deformations and high strain rates, the elastic component of the deformation is small  with respect to the inelastic one, it can be neglected and   a rigid-viscoplastic approach can be adopted (see for instance,\cite{ref50:friderikos2020simulation,lebtome93,kok02,smiri:tel-05010824,Salman2021-ts}). This means that  we  neglect  the elastic lattice strain $\Ub^e$ by supposing that  $\Ub^e \approx  \Ib $.  This leads to the following decomposition of the deformation gradient $\Fb$ (see \cite{kok02}): 
\begin{equation}
		\Fb=\Rb\Pb.
	\label{RP}
\end{equation}
Such a hypothesis is valid since during forming or other industrial processes, the elastic component of deformation is negligibly small (typically 10$^{-3}$) in comparison to the plastic component (typically $>$10$^{-1}$). It also to be noted that once the elastic/plastic transition is over the stress evolution in the grains is controlled by plastic relaxation, see  \cite{lebtome93}. 
Since elastic effects are neglected, we have 
$	\bbs_s = \Rb \bbs_s^0, \quad   \bm_s = \Rb \bm_s^0$
%
and 
$
	\bbs_s\otimes \bm_s = \Rb\left(\bbs_s^0 \otimes \bm_s^0\right)\Rb^T
$.

We seek  to express the lattice evolution equations only in terms of vector and tensor fields associated with the current configuration. Let  $\bu=\bu(t,x)$, the Eulerian velocity field,  $\Lb$ the velocity gradient,  $\D$ the rate of deformation,  and $\W$ the spin tensor
\begin{equation}\label{DW}
	\Lb=\Lb(\bu)=\nabla \bu, \quad  \D=\D(\bu)=(\nabla \bu)^{symm}, \quad \W=\W(\bu) =(\nabla \bu)^{skew}.
\end{equation}
    If we  denote by  
$
	\M_s= \left(\bbs_s\otimes\bm_s\right)^{symm}, \quad \Qb_s=\left(\bbs_s\otimes\bm_s\right)^{skew}
$ 
then,  using $\Lb= \dot{\Fb}\Fb^{-1}= \dot{\Rb}\;\Rb^{T} + \Rb\dot{\Pb}\Pb^{-1}\;\Rb^{T}$  and  (\ref{flowrulenoi}), the rate of deformation $\D$   can be  written as
\begin{equation}\label{D}
	\D = \sum_{s=1}^N \dot{\gamma}^{s}\M_s. 
\end{equation}
Taking the  anti-symmetric  part of  $\Lb$,  we  obtain that the spin tensor is $\W=\dot{\Rb}\;\Rb^{T}+\sum_{s=1} ^N\dot{\gamma}^{s} \Qb_s$  and a differential equation for the rotation tensor $\Rb$:
\begin{equation}\label{Rp}
	\dot{\Rb}=(\W-\sum_{s=1} ^N\dot{\gamma}^{s} \Qb_s)\Rb. 
\end{equation} 
The evolution equations (\ref{Rp})  describe the evolution of the lattice in terms of vector and tensor fields associated with the current configuration.

\subsubsection{In-plane  deformation} 
\label{sec:fccplane}
\begin{figure}[ht]
	\begin{center} 
		\includegraphics[scale=0.25]{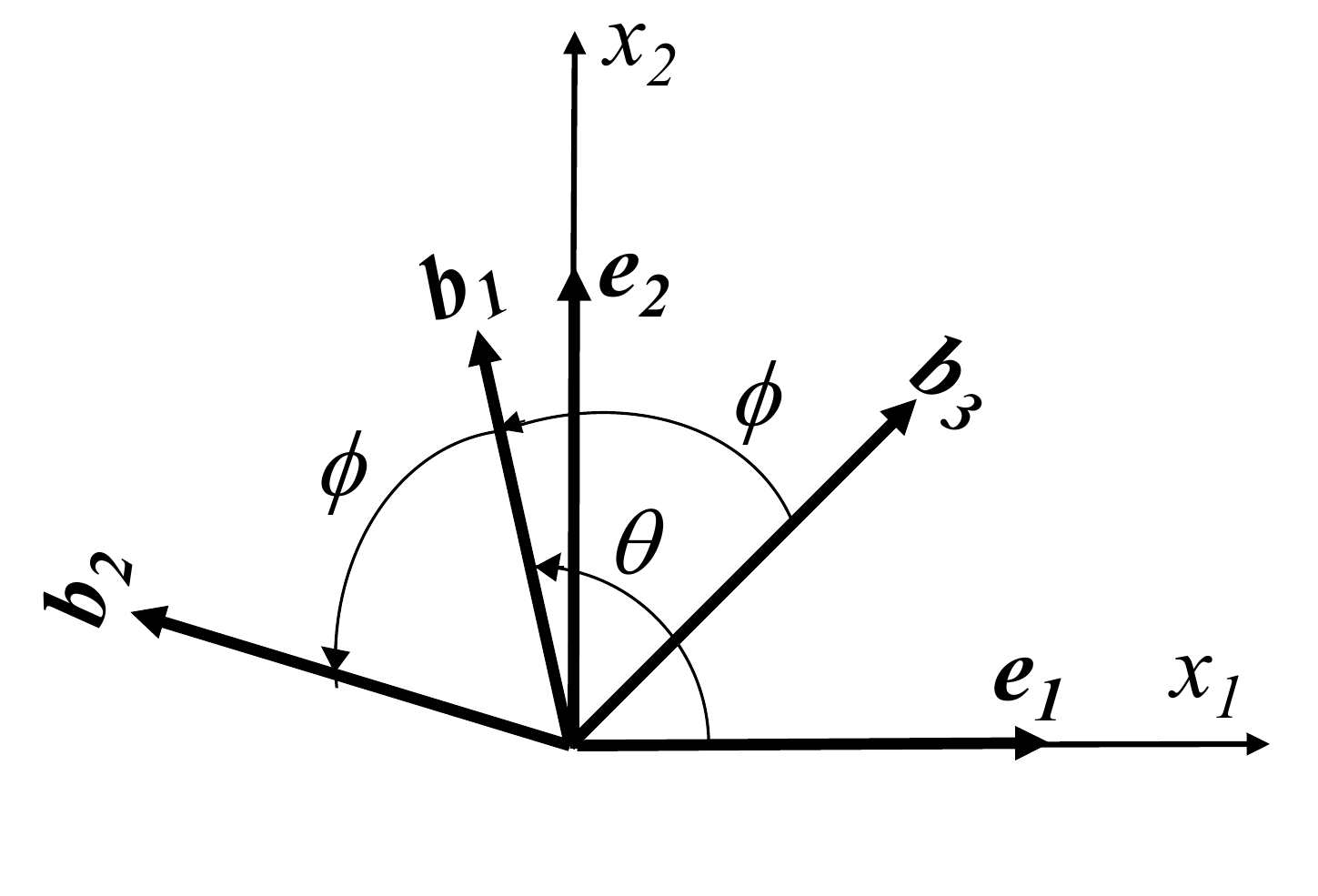}
	\end{center}
	\vspace{-.9cm}
	\caption{Two dimensional model with three slip systems.} \label{3slip}
\end{figure}

The 3D system with a large number of slip systems  $N$  is too difficult to be analyzed from
a theoretical or physical point of view. That is why, in many situations, a simplification could be useful for a better understanding of the complex phenomena which occur in crystal deformation. We will introduce here a two-dimensional model with $N=3$ slip systems. In this case let us denote with $\phi$ and  $-\phi$   the angles  formed by slip system $r=1$ with the other two systems $r=2,3$ and let  $\theta$ be the  angle formed by the slip system $1$ with the $Ox_1$ axis (see Fig. \ref{3slip}).   The three composite in-plane slip systems  $ \bbs_1, \bbs_2,  \bbs_3$ are  specified  by  the angles  $\theta,  \theta+\phi,  \theta-\phi$.
\label{sec:TOYMODEL}
The main simplification  for the 2-D problem comes from the  latice rotation  $\Rb$ which is now a rotation  $\Rb(\theta, \be_3)$ with angle $\theta$ along $Ox_3$ axis and  we have 
$
	\Qb_r=\dfrac{1}{2}\left((1,0)\otimes(0,1) -  (0,1)\otimes(1,0)\right)
$ 
for all $r=1,2,3$ and  (\ref{Rp}) has  a much more simpler form 
\begin{equation}\label{theta}
	\dot{\theta} = \frac{\partial \theta}{\partial t} + \bu\cdot \nabla \theta= \frac{1}{2}\left( \sum_{r=1}^{3}\dot{\gamma}^{r} -(\frac{ \partial v_1}{ \partial x_2}-
	\frac{ \partial v_2}{ \partial x_1})\right),
\end{equation}
where $\bu=(v_1,v_2)$ is the Eulerian velocity field. We have in mind two situations  where this model is physically sound : 

\textit{i) In-plane  deformation of a FCC crystals :} 
Rice \cite{ric71} showed that certain pairs of the  three-dimensional systems that are potentially active need to combine in order to achieve plane-strain deformation.   For a F.C.C. crystal, with 12 potentially active slip systems,    we consider  $Ox_3$ axis to be parallel to  $ [1 1 0] $ in the  crystal basis, which  means that   the plane-strain  plane $(Ox_1x_2)$ is   the plane  $[\bar{1}10]-[001]$.   Some geometrical constraints  (see \cite{Cazacu2010-hi})  must be satisfied such that  $N=3$ pairs of   composite  systems will give deformation in the plane $(Ox_1x_2)$.  The angle $\phi$  between  the slipping system $1$ and $2$ is $\phi=\arctan(\sqrt{2}) \approx 54.7^\circ.$ Since there are  some scalar factors between the first two components  of the in-plane systems and the 3-dimensional ones given by  
$q_1=\frac{1}{\sqrt{3}}, \quad q_2=q_3= \frac{\sqrt{3}}{2},$
the 2-D composite slipping rate $\dot{\gamma}^r$ corresponds to the  3-D  slipping rate $\dot{\gamma}^k$  multiplied by $2q_r$.  As it follows from \cite{Cazacu2010-hi}  the  2-D yield limit   $\tau_0^r$ corresponds to the  3-D yield limit   $\tau_0^k$  divided by $q_r$ while the  2-D   viscosity   $\eta^r$ corresponds to the  3-D yield limit   $\eta^k$  divided by $q_r^2$.

{\em ii) Slip in the basal plane of a hexagonal crystal.}  Alternatively, this situation corresponds to hexagonal close-packed (HCP) crystals (such as Ti, Mg, Zr, etc) under plane strain, with the plane of deformation aligned with the basal plane (0001) of the hexagonal lattice. This situation has been considered experimentally in \cite{crepin1996}. In this situation, the plastic strain is mainly accommodated by the three prismatic slip systems, i.e. the $(10\bar{1}0)\langle1\bar{2}10\rangle$ slip family. Each of those slip systems can act independently from the two others, as its strain is a plane strain. One might remark that combinations of other slip systems of HCP crystals from the basal and the pyramidal slip families could also lead to plane strain, but it would require significantly more energy, and therefore never occur. Thus, those slip families are not considered in this work. The prismatic slip systems are symmetric, leading to  $\phi=\pi/3= 60^\circ.$

\subsection{Saturation dislocations of the 2D model}  

\label{Sat2D}

Let us analyze here more in details the  simplified model to see how the  methodology, previously described,  is applied to examine the stability of dislocation density evolution given by the KM model of a 2D model with three active slip systems.

 For all numerical computations presented in this section we have used  the 2D in-plane  FCC model (i.e. $\phi=54.7^\circ$) with he reference value \( \tau_0 \), along with all other material parameters used in the simulations, are provided in Tables~\ref{tab:FCC_asp_Ni} and~\ref{tab:FCC_Ni}. This yields to \( \tau_0^1 = \tau_0/q_1=19.052 \) MPa and \( \tau_0^2 = \tau_0^3 = \tau_0/q_2=12.705 \) MPa. A uniform initial dislocation density is chosen across all slip systems.

\begin{itemize}
 	\item (i)  $\rho_{0}^{s} = \rho_{0}=9 \times 10^{13} \; m^{-2} \; \mbox{for all} \; s=1,2,3,$
 	\item  (ii)  $\rho_{0}^{s} = \rho_{0}=7.5 \times 10^{12}\;  m^{-2} \; \mbox{for all} \; s=1,2,3$. 
\end{itemize}	

\subsubsection{Attractors of self interaction dislocations} 
In the context of self-interaction within the simplified model, the saturation of dislocation density and shear yield limit are determined using equation~(\ref{KM-self}). Notably, the saturation dislocation density is uniform across the three slip systems and independent of specimen size, with the value 
\(
\rho_{sat}^{s} = 3.1169 \times 10^{13} \; \text{m}^{-2}, \quad \text{for all } s = 1, 2, 3
\). Since the initial shear yield limit are different the saturation shear yield limit $\tau_{c,sat}^{s}$ is not uniform, exhibiting system-dependent values: \(
\tau_{c,sat}^{1} = 28.13 \; \text{MPa}, \; \tau_{c,sat}^{2} = \tau_{c,sat}^{3} = 21.783 \; \text{MPa}.
\)

\begin{figure}
	\centering
	\begin{minipage}{0.48\textwidth}
		\centering
		\includegraphics[scale=0.3]{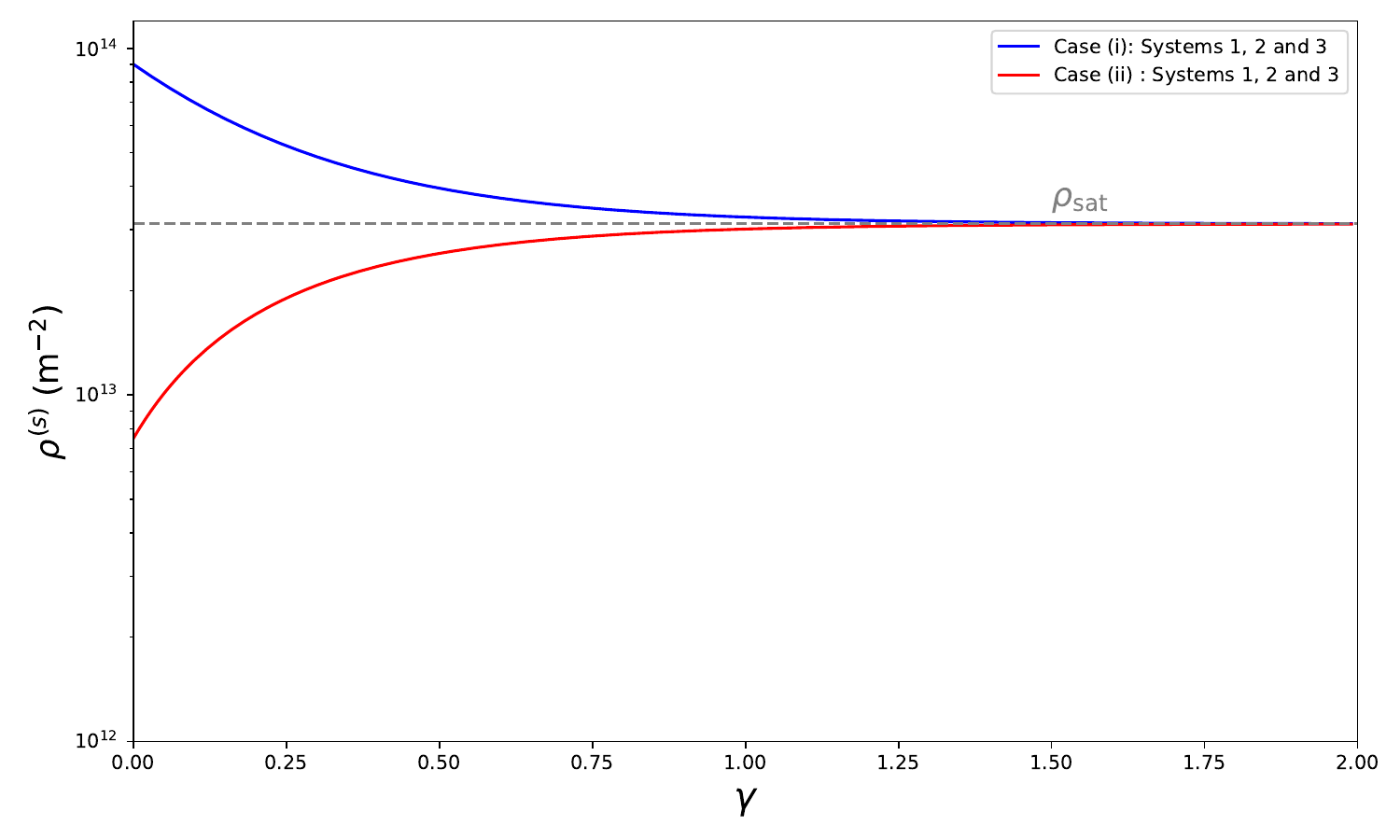}
	\end{minipage}
	\hfill
	\begin{minipage}{0.48\textwidth}
		\centering
		\includegraphics[scale=0.3]{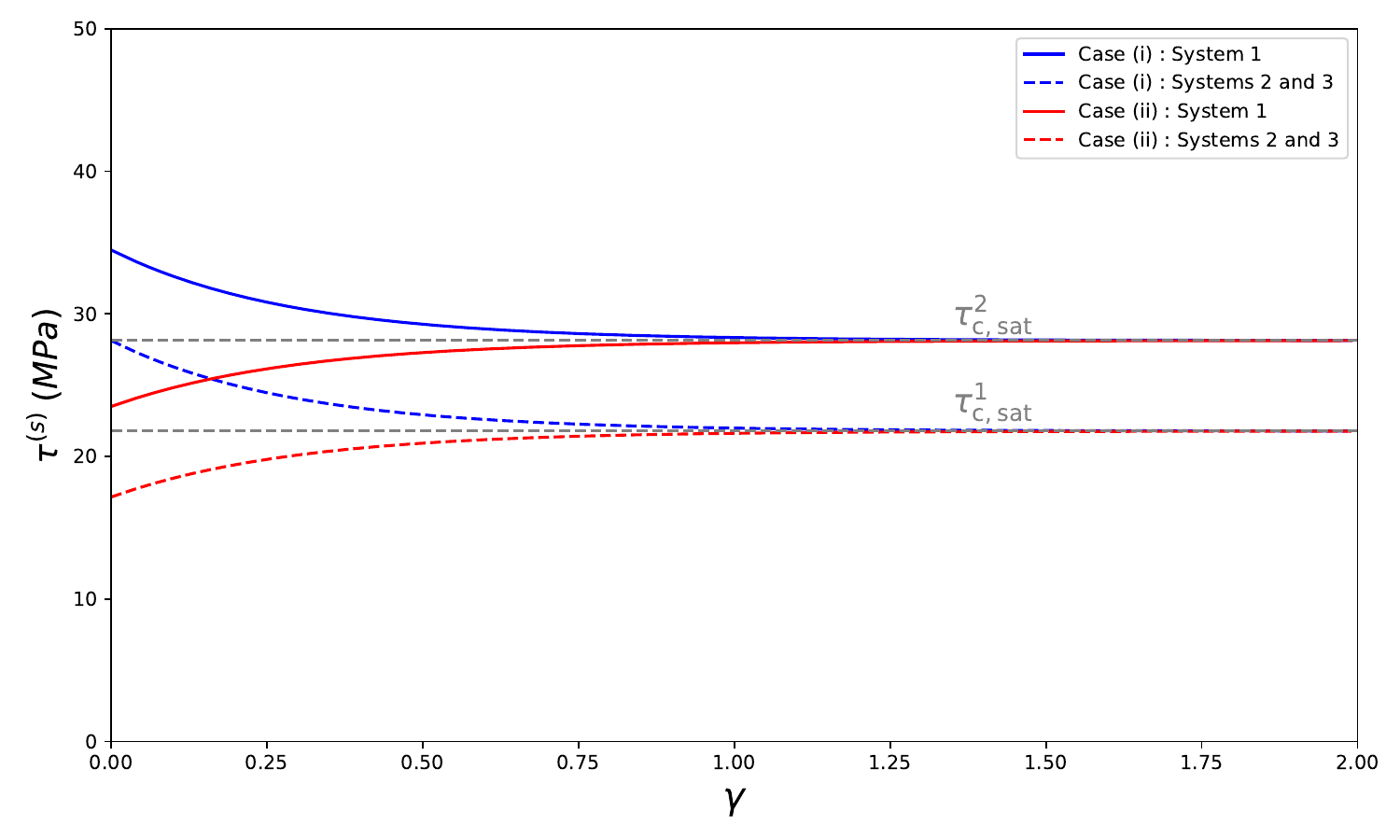}
	\end{minipage}

	\caption{Evolution of the self-interaction 2-D model for two different  choices of initial dislocation densities $\rho_{0}^s$, corresponding to case (i) 
		(in blue) and case (ii)  
		 in (red). Left: dislocation density  (from KM model in m$^{-2}$) over slip $\gamma  \to \rho^s(\gamma)$. Right: shear yield strength (in MPa) over slip $\gamma  \to \tau^s_c(\gamma)$ with $s=1$ in solid line and $s=2,3$ in dashed line.}
	\label{DD_tau_self}
\end{figure}

This distinction becomes evident in the two scenarios illustrated in Figure~\ref{DD_tau_self}. In case (i), where the initial dislocation density of the 3 systems exceeds the saturation value, leading to softening: as $\rho^{s} \rightarrow \rho_{sat}^{s}$, the CRSS $\tau^{s}(t)$ declines toward its system-specific saturation value—$\tau_{c,sat}^{1}$ for system 1 and $\tau_{c,sat}^{2}$ for systems 2 and 3. (Although the model permits at most two simultaneously active slip systems, all three are shown as active in the figure for illustrative purposes.). 

Conversely, in case (ii), representing a larger specimen, the system experiences hardening due to relatively low initial dislocation densities compared to the DD saturation $\rho_{sat}^{s}$. Both $\rho^{s}$ and $\tau^{s}(t)$ increase: $\rho^{s}$ tends toward the uniform saturation value $\rho_{sat}^{s}$, while $\tau^{s}(t)$ approaches its system-specific saturation value—$\tau_{c,sat}^{1}$ or $\tau_{c,sat}^{2}$ depending on the system in question.

\bigskip 

\subsubsection{Attractors of cross-interaction dislocations}
	\label{Att2D}

In the analysis of cross-interactions, we utilize equation (\ref{SatKM}) to compute the stationary points. This is a nonlinear system which does not have, generally, analytical solutions. 
 However in our simple model  assumption  (\ref{Sa})    is valid  and we can use  (\ref{Rosatcross}). We have considered  the dislocation interaction matrix associated with the three active slip systems  as follows: $a_{11}=a_{22}=a_{33}=a_{0}^{self}$ and  
 $a_{12}=a_{13}=\beta=a_{cop}^{coplanar}$ and $a_{23}=\zeta=a_{cop}^{coplanar}$.   Concerning the choice of active and inactive systems two cases, inspired from the FE computations  of micro-pillars compression, will  be considered in the next: 
 \begin{itemize}
 	\item (a)  $A_c=\{1\}, A_i=\{2,3\}$
 	\item  (b)  $A_c=\{1,2\}, A_i=\{3\}$
 \end{itemize}	

\bigskip 
 	
 Case (a) has  only one active system $s=1$, hence  $a_c=a_0$ and $\rho_{0,i}=\rho_0(\beta+\zeta)$ and we can compute the saturation values  from (\ref{Rosatcross}) as: 
\begin{equation} \label{Rosatcross2D}
 \rho_{sat}^1=\frac{a_0 + \sqrt{a_0^2+4(2y_{c}k)^2\rho_0(\beta+\zeta)}}
 	{2(2y_ck)^2},\quad 	\tau_{c,sat}^{s}=\tau_0^s+\alpha \mu b \sqrt{\rho_{sat}^1  d^{s1} + (d^{s2}+d^{s3}) \rho_0}.
 \end{equation} 

\begin{figure}
	\centering
	\begin{minipage}{0.48\textwidth}
		\centering

		\includegraphics[scale=0.3]{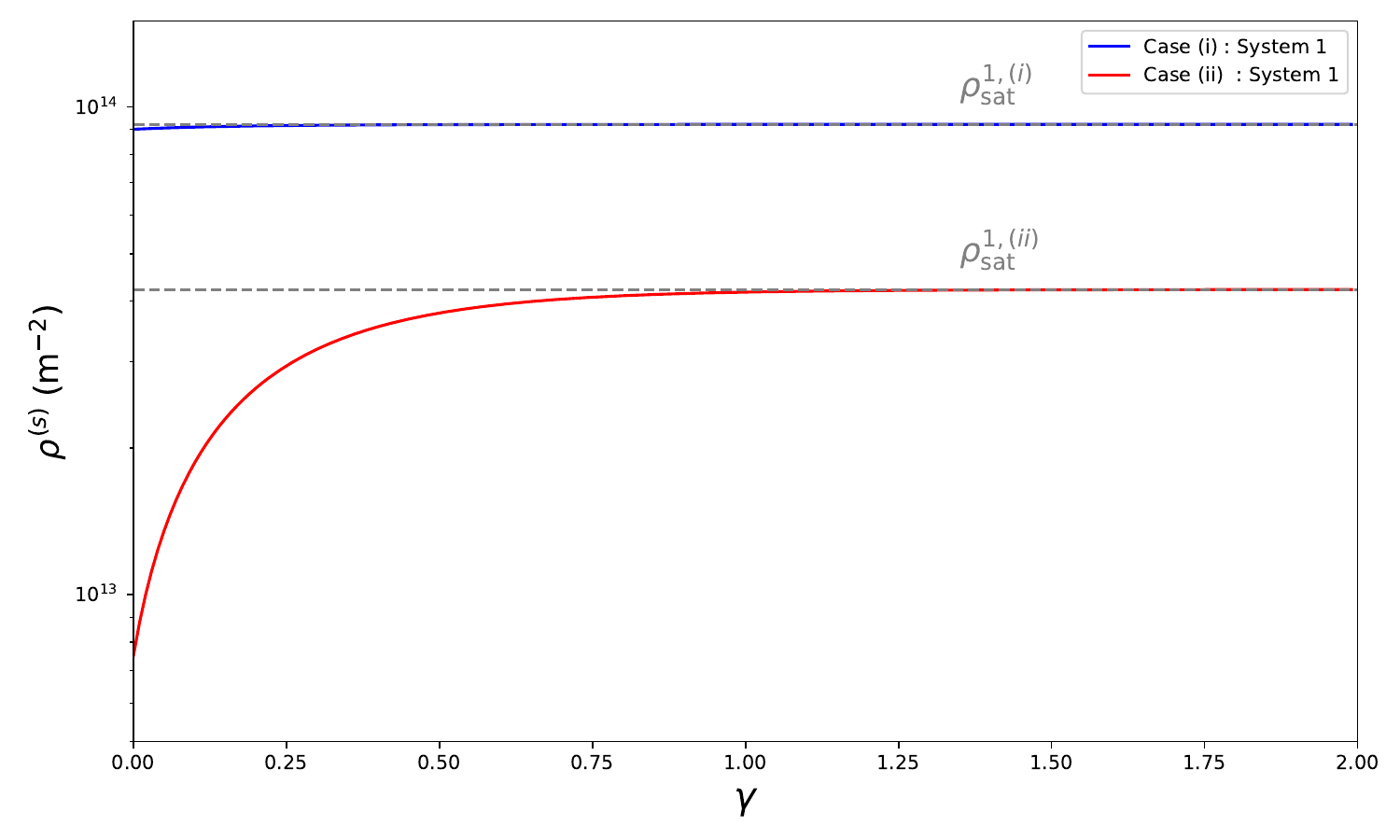}
	\end{minipage}
	\hfill
	\begin{minipage}{0.48\textwidth}
		\centering
		\includegraphics[scale=0.3]{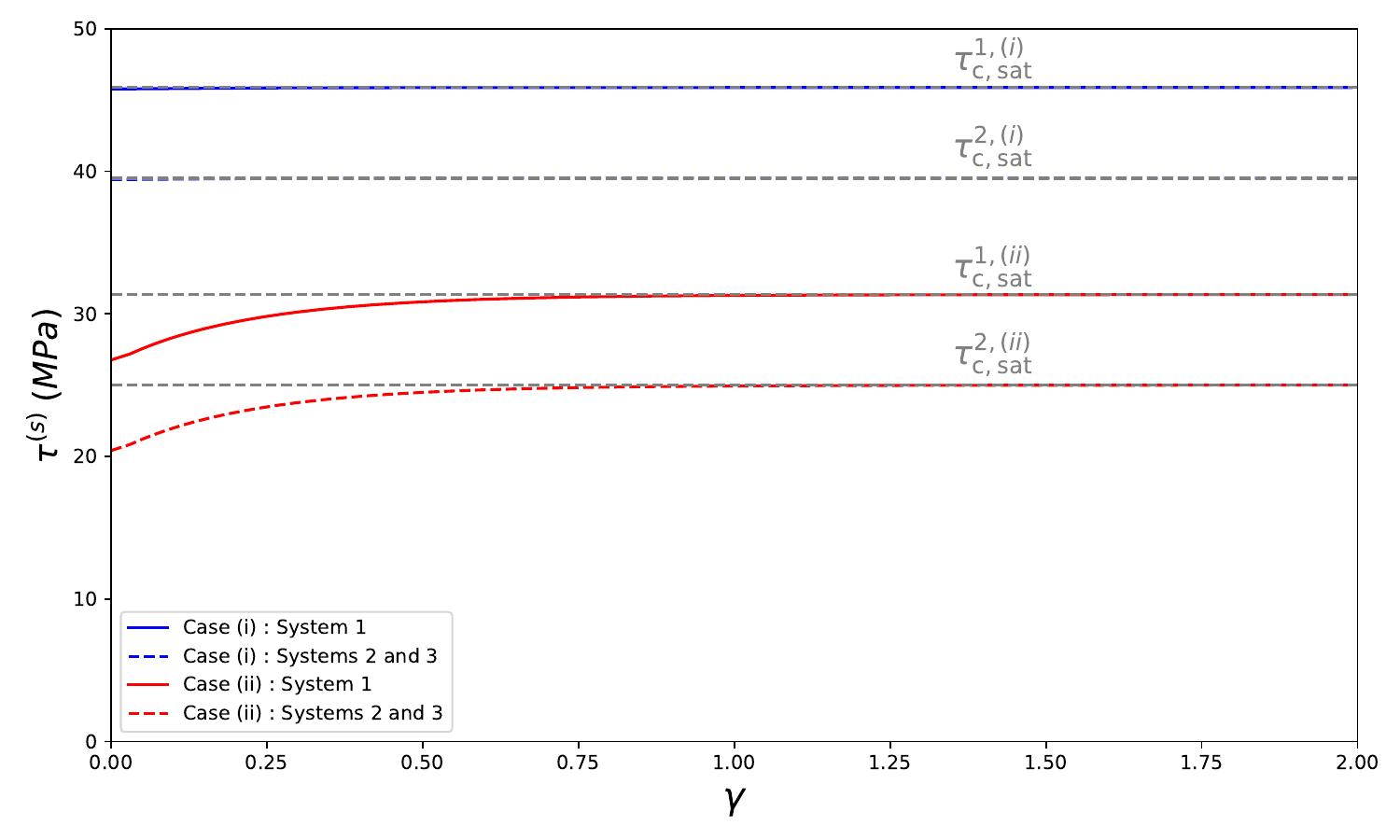}
	\end{minipage}
	\caption{Evolution of the cross-interaction simplified model (\ref{Rosatcross2D})  in case (a) for two different initial choices of initial dislocation densities $\rho_{0}$, corresponding to case (i)  (in blue) and  case (ii) (in red).  Left: dislocation density (in m$^{-2}$)  over slip $\gamma  \to \rho^1(\gamma)$. Right: shear yield strength (in MPa) over slip $\gamma  \to \tau^s_c(\gamma)$, shown for slip system $s=1$ (solid line), and for $s=2$ and $s=3$ (dashed line; both curves are superimposed).}
	\label{fig:DD_tau_UC1}
\end{figure}

Figure~\ref{fig:DD_tau_UC1} illustrates the response of the  cross-interaction model (equation~{\color{sepia}\ref{Rosatcross2D}}), showing the evolution of both dislocation density and shear yield strength with increasing slip, for two different initial dislocation densities. This allows for a direct comparison between case (i) in blue and case (ii) in red. The dislocation density evolution is shown only for the active slip system \( s=1 \). For the inactive systems \( s=2 \) and \( s=3 \), dislocation densities remain constant throughout the deformation, resulting in flat lines that are not plotted here for clarity. 
Despite the inactivity of these systems, the corresponding critical resolved shear stresses (CRSS) still evolve due to cross-interaction effects; hence, CRSS curves are shown for all three systems. For system \( s=1 \), we observe that the dislocation density evolves toward a saturation value that depends on the initial configuration. This is a key distinction from the self-interaction case, where the saturation dislocation density \( \rho_{sat}^s \) are independent of the initial state. The cross-interaction model thus captures a more sensitive dependence of the hardening behavior on the initial micro-structural conditions.
 The dislocation density saturates at 
	\( \rho_{sat}^{1,(i)}= 9.26 \times 10^{13}~\text{m}^{-2} \) for case (i)  and 
    \( \rho_{sat}^{1,(ii)} = 6.58 \times 10^{13}~\text{m}^{-2} \) for  case (ii). 
	In contrast, the dislocation densities in systems \( s = 2 \) and \( s = 3 \) remain constant during loading, and these constant values do not result from a saturation process.

 	It is also worth noting that several distinct saturation values are observed for the CRSS: two for system \( s = 1 \), corresponding to each initial configuration: $\tau_{c,sat}^{1,(i)}=45.87$MPa for case (i)  and $\tau_{c,sat}^{1,(ii)}=31.35$MPa for case (ii),   two for the inactive systems \( s = 2 \) and \( s = 3 \), which also correspond to the two cases  $\tau_{c,sat}^{2,(i)}=\tau_{c,sat}^{3,(i)}=39.51\,$MPa for case (i) and $\tau_{c,sat}^{2,(ii)}=\tau_{c,sat}^{3,(ii)}=25.00\,$MPa for case (ii). 
 	The CRSS evolution in these inactive systems is instead driven solely by cross-interactions with the evolving dislocation density of system~1. This quasi-stationary dislocation structure across the three systems, maintained as slip increases, leads to a plateau-like mechanical response. This contrasts with the behavior observed in the self-interaction model, where softening is more pronounced due to the absence of cross-interaction effects.

\bigskip 

Case (b) has  two  active system $s=1, 2$, hence  $a_c=a_0+\beta$ and $\rho_{0,i}=\rho_0\zeta$ and we can compute the saturation values  from (\ref{Rosatcross}) as: 
$$
	\rho_{sat}^1=\rho_{sat}^2=\frac{a_0+ \beta+ \sqrt{(a_0+\beta)^2+4(2y_{c}k)^2\rho_0\zeta}}
	{2(2y_ck)^2},$$
	$$ 	\tau_{c,sat}^{s}=\tau_0^s+\alpha \mu b \sqrt{\rho_{sat}^1  (d^{s1} +d^{s2})+ d^{s3} \rho_0}.
$$

\subsection{Numerical simulations setup}  

\begin{figure}[!hbt]
\begin{center}
	\includegraphics[scale=0.3]{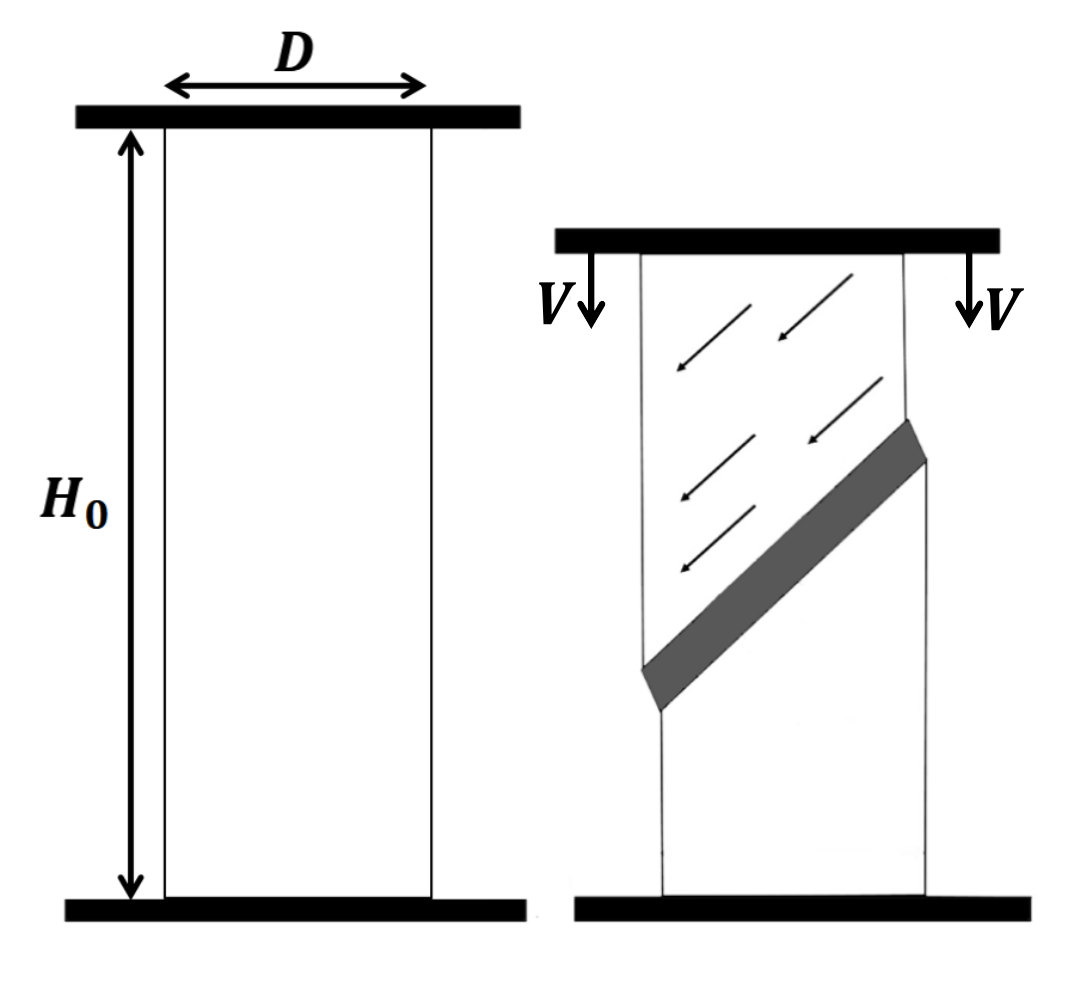}
\end{center}
\caption{A schematic representation of micro-pillar compression and the formation of a shear band (in grey).} \label{shematiccompression}
\end{figure}

The compression test methodology is shown schematically in Figure \ref{shematiccompression}. The  aspect ration between the micro-pillar initial  high  $H_0$  and its the diameter  $D$ was chosen  to  be {\color{sepia} 2.5}. The setting mimics the compression experiment commonly conducted on nano and micro scale samples. The (top) surface of the sample is in unilateral frictional contact (friction coefficient $\mu_f=0.2$) with a rigid plate which moves with a small constant velocity $-V$ on $y$-axis, the plate movement continues until the sample's final height is reduced by $25\%$ from the initial \textcolor{black}{height}, i.e.  the   engineering  strain $\eps^{engn}=(H_0-H)/H_0$  is  increasing up to  $\eps^{engn}_{final}=0.25$. The other parts of the pillar which is not in contact with the upper plate are traction free whereas on the bottom the sample is fixed.   

In all the simulations, the global rate of deformation is kept low at 0.5\,s\(^{-1}\), allowing the loading to be considered quasi-static. For the material coefficients of Nickel, all parameter values—except for the mass density \( \rho^{\text{mass}} = 8900 \,\text{kg/m}^3 \)—are provided in Tables~\ref{tab:FCC_Ni} and~\ref{tab:FCC_asp_Ni}. These include the shear modulus, Burgers vector, the KM dislocation density evolution model parameters, as well as the slip system interaction matrix.
 
\textit{Initial dislocation densities.} 
We used two different initial dislocation densities, denoted as (i) and (ii), and described in Section~\ref{Sat2D}.

\textit{Initial orientation of the mono-crystal} was chosen to be  $\theta^0 = 65^{\circ}$.

\textit{Viscosity.} The viscosity was chosen to be as small as possible  (less  than $1\%$  of $\tau_{0} H/V$) to ensure the convergence of the numerical scheme while maintaining the rate-independence of the mechanical model.

\textit{Time step.} Since we are using an implicit numerical scheme, the chosen time step is relatively large and corresponds to a deformation of $0.1\%$ between two time steps. As a result, the computational cost is low, allowing the simulations to be performed on a personal computer.

\begin{figure}[!hbt]
	\center
	\includegraphics[width=3.5cm,keepaspectratio]{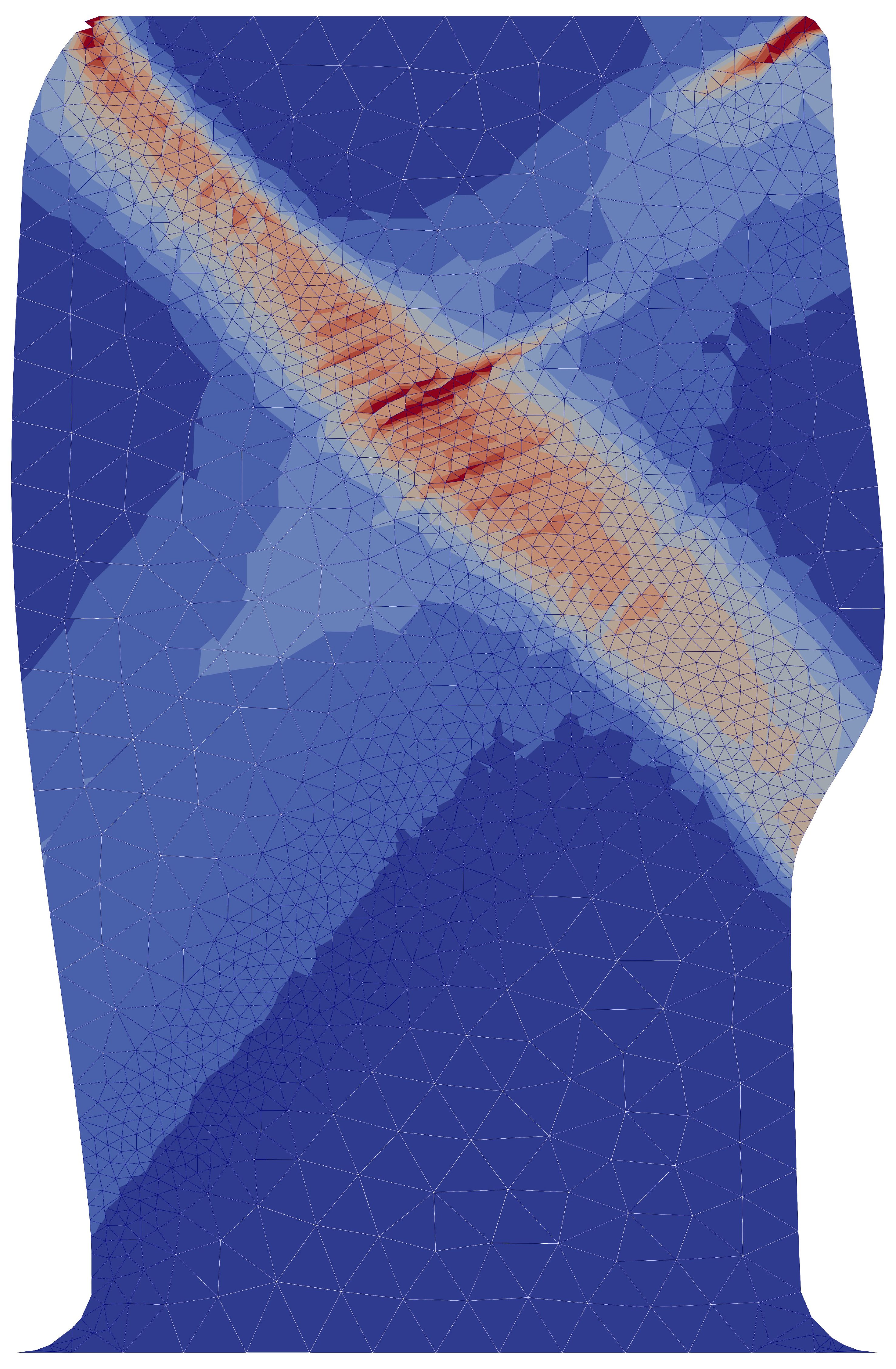}
	\includegraphics[width=3.5cm,keepaspectratio]{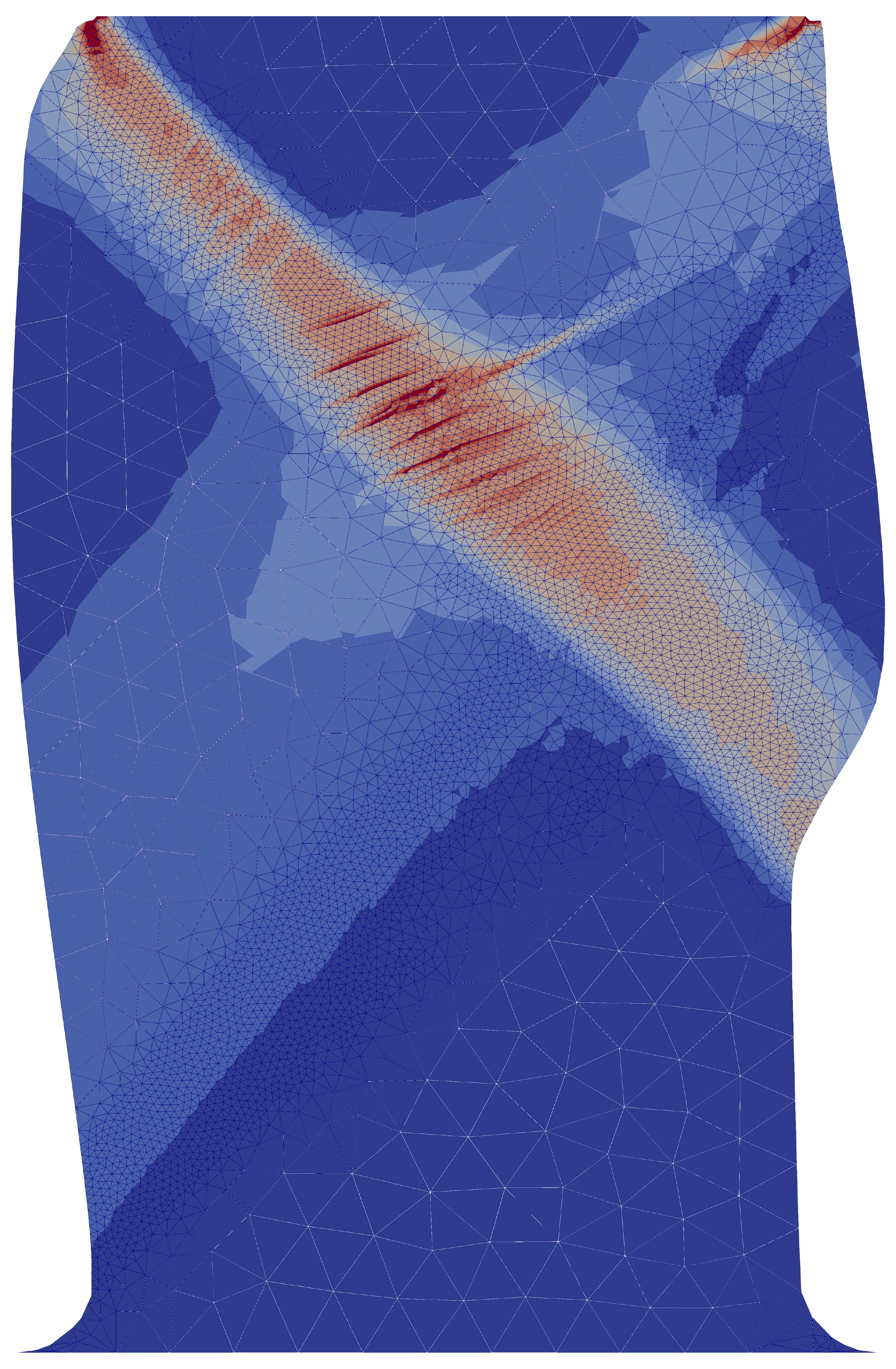}
	\includegraphics[width=3.5cm,keepaspectratio]{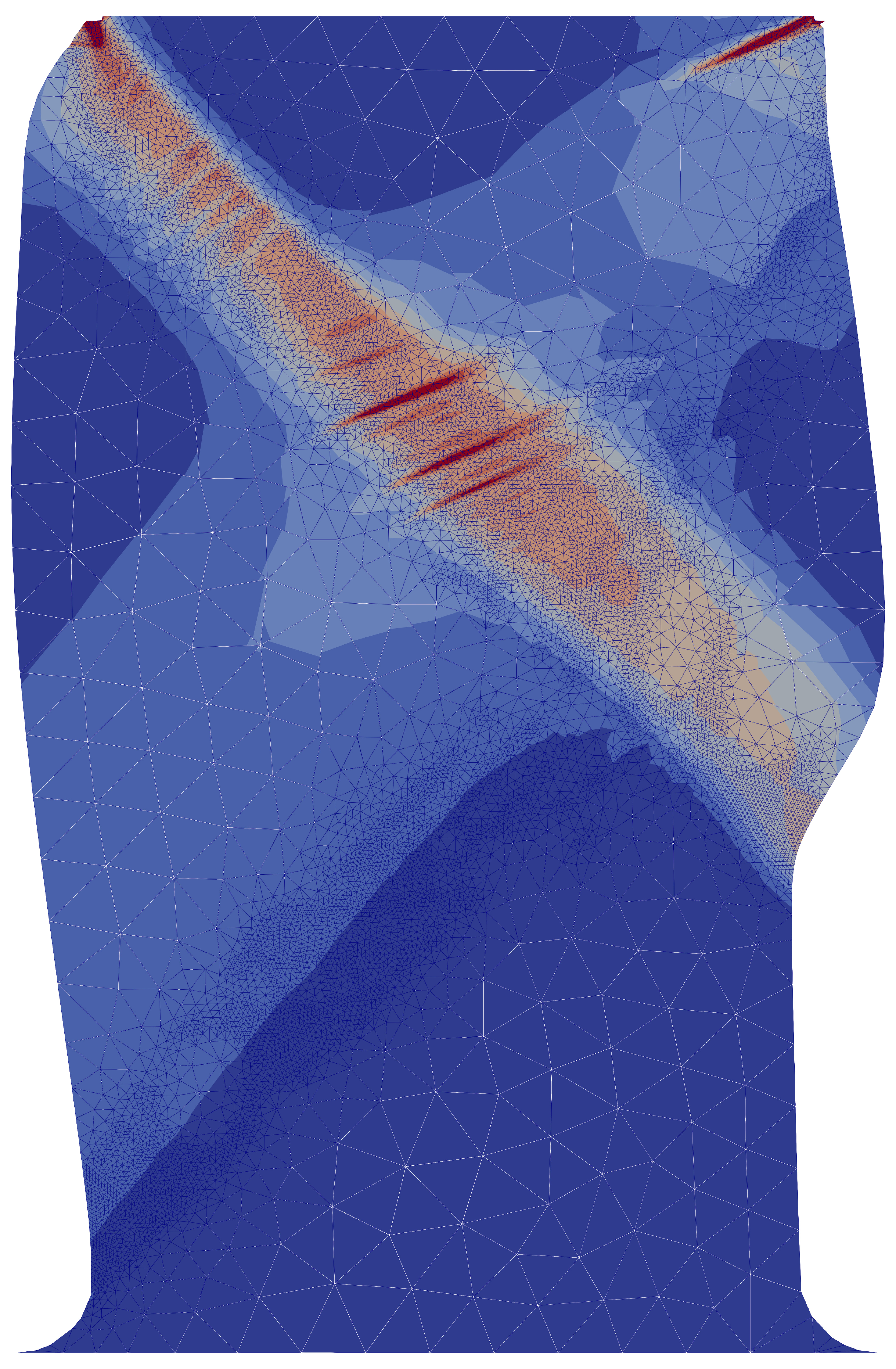}
	\includegraphics[width=1cm,keepaspectratio]{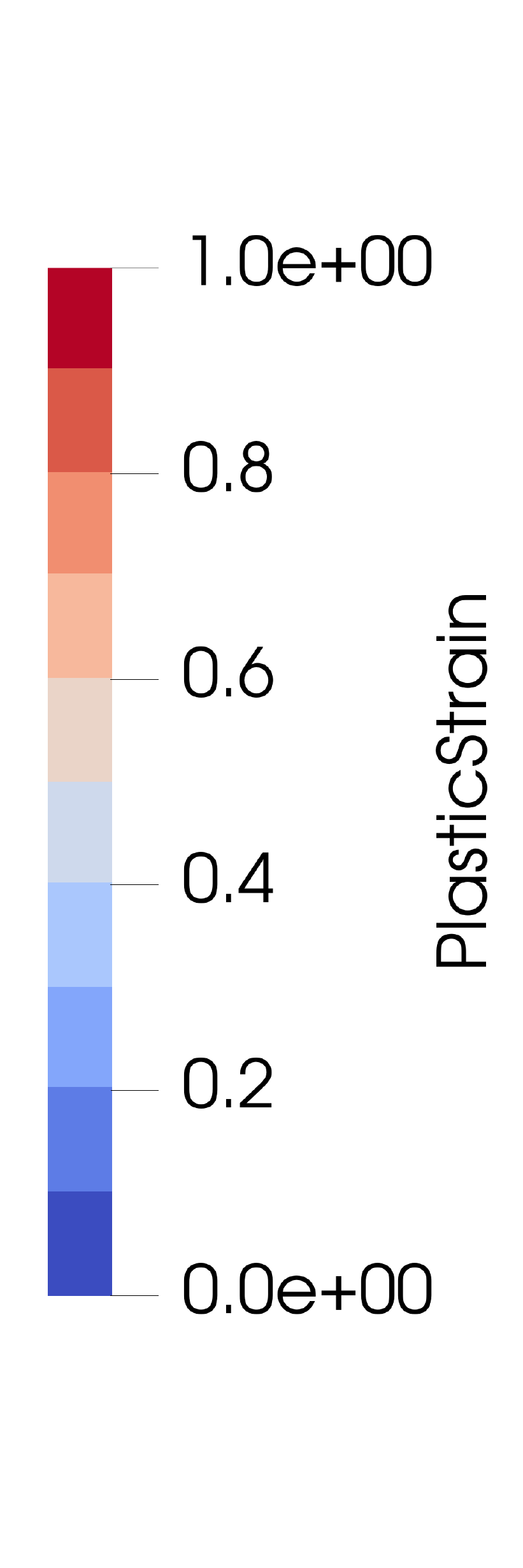}
	\caption{The computed distribution of cumulative plastic strain $\epsilon_p$  at $\eps^{engn}=15\%$ in case (i) for three different meshes with different ratio between the sizes of the fine and coarse meshes:  $1/4$ (left);  $1/8$ (middle);  $1/16$(right).} \label{differentmeshratio}
\end{figure}

\textit{Mesh.} To capture the shear bands, we used an adaptive mesh technique based on the strain rate norm 
$\dot{\epsilon}_p = \vert \D \vert,$ 
where $\epsilon_p$ is the cumulative plastic strain. This means that regions with a higher slip rate have a finer mesh, while regions outside have a coarser mesh. The ratio between the fine and coarse mesh sizes was $1/8$.  Figure \ref{differentmeshratio}, shows the distribution of cumulative plastic strain $\epsilon_p$ at $\eps^{engn}=15\%$  for three mesh refinement ratios. We found that for the ratio $1/8$ is small enough to assure that the numerical results are not mesh-dependent.

\subsection {Numerical scheme used for the simulations}

	The main goal of this section is to recall the 
	Eulerian numerical strategy proposed in \cite{CI09}  (see also \cite{Smiri:2024aa}) for  the   rigid-visco-plastic  crystal   model used in this work.  	Time implicit (backward) Euler scheme  is used  for the field equations, which 
gives  a set of nonlinear equations for the velocities $\bu$ and lattice orientation $\theta$.  
At each  iteration in time,  an iterative  algorithm is used in order to solve these nonlinear equations. Specifically, the variational  formulation  for the velocity field is discretized using the finite element method, while a  Galerkin discontinuous method with an upwind choice of the flux is adopted for solving the hyperbolic equation that describes the evolution of the lattice orientation. One of the advantages of the proposed numerical strategy is that it does not require the consideration of elastic deformation since it is not based on an elastic predictor/plastic corrector scheme (e.g. \cite{SH}).

It is to be noted that in the case of the proposed rigid-viscoplastic model  (\ref{flow_rule}), numerical difficulties arise from the non-differentiability of the viscoplastic terms.  That means that one cannot  use  finite element techniques developed for Navier-Stokes fluids.
 To overcome these difficulties the iterative  decomposition-coordination formulation coupled with the augmented Lagrangian method of    \cite{GlLT}  
 was  modified. 
The reason for this modification  is that in the used crystal model, there is not co-axiality between the stress deviator and the rate of deformation as it is the case in the von-Mises (Bingham)  model used in 
\cite{GlLT}.    
This type of algorithm permits also to solve alternatively,  at each iteration,  the equations  for the velocity field and   for the unit vectors that define the lattice orientation.  

For vanishing   viscosity,  the adopted  visco-plastic model  contains as a limit case the inviscid Schmid law.  Even that Schmid model  is very stiff,  for small viscosities (as for metals)  and moderate strain rates the iterative decomposition coordination formulation coupled with the augmented Lagrangian method  works very well and no instabilities are presents.  

To  include frictional effects in the algorithm, we first regularize the friction acting on the  upper part of the micro-pillar. This was  done (as in  \cite{ionescu2010onset,ionescu2013augmented}) by introducing a small frictional viscosity $\eta_f$  in the Coulomb friction law    as
\begin{equation} \label{fricreg}
	[\bu_{T}]= -\frac{1}{\eta_f} \left[1-\frac{\mu_f [-\sigma_{n}]_+}{\vert \bs_T \vert}\right]_+ \bs_T,
\end{equation}
where $[\bu_{T}]$ is the relative tangential velocity, $\sigma_{n}, \bs_T$  are the normal and tangential stresses, $\mu_f$ is the frictional coefficient and $[ x] _+=(x+\vert x\vert)/2$. Note that using this regularization, the friction law has the same mathematical structure as the viscoplastic constitutive equation  and we can use the same iterative  decomposition-coordination formulation.  

\bigskip
If the Eulerian  domain $\DD$  has time variations  then the above algorithm has been  adapted  to an ALE (Arbitrary Eulerian-Lagrangian) description of the crystal evolution.  In this case is more convenient to have the same finite element Galerkin discontinuous meshes. This avoid the interpolation of the lattice orientation  on the deformed mesh.   As a matter of fact, the numerical algorithm proposed here deals only with a Stokes-type problem at each time step and the  implementation of the Navier-Stokes equations in an ALE formulation is rather standard  (see for instance
\cite{Jafari2016,Duarte2004}). 

\subsection{Self interaction simulations} 
\begin{figure}[!hbt]
	\center
	\includegraphics[width=4cm, keepaspectratio]{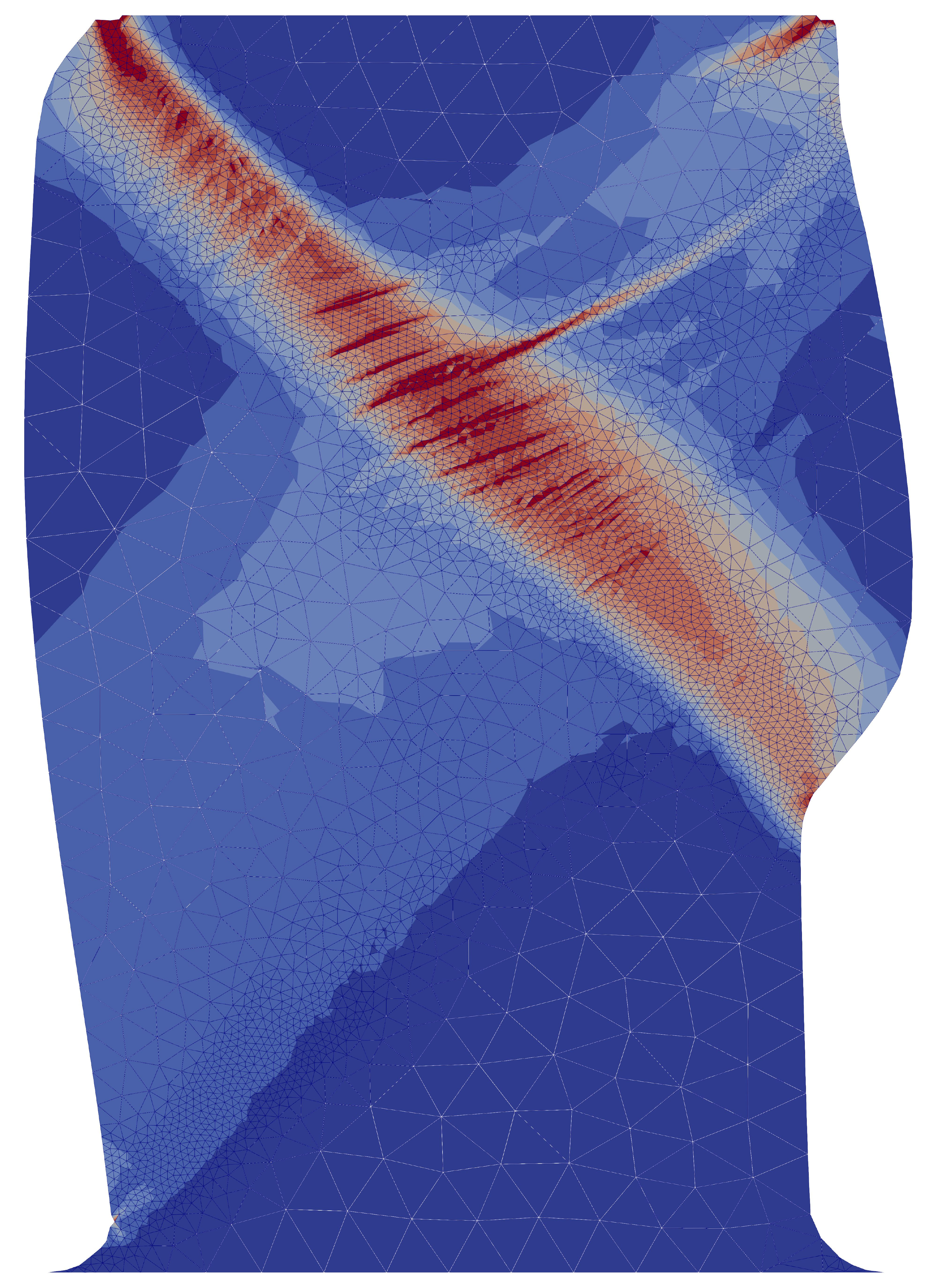} \quad \quad 
	\includegraphics[width=4cm, keepaspectratio]{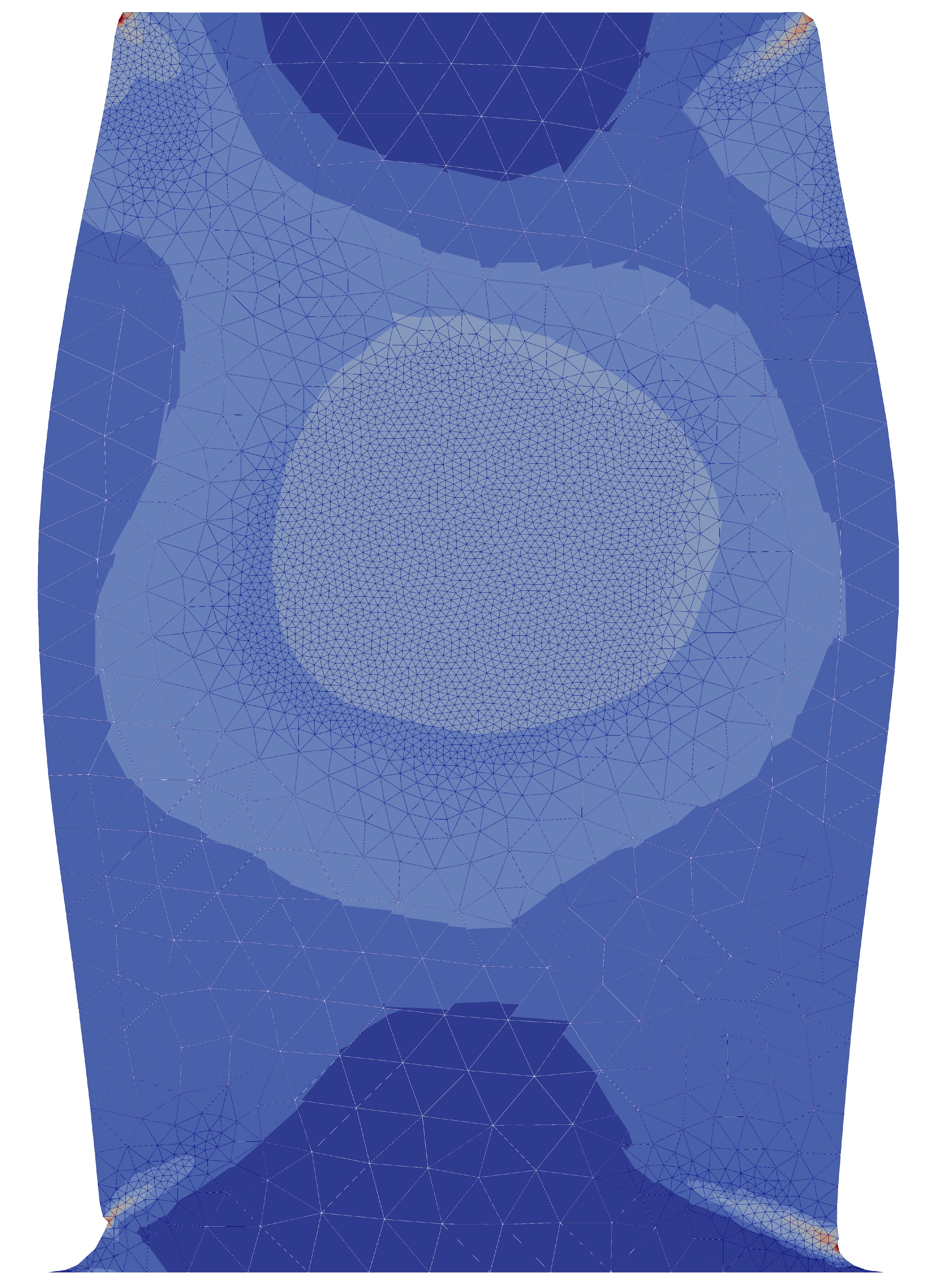}
	\includegraphics[width=1.5cm,keepaspectratio]{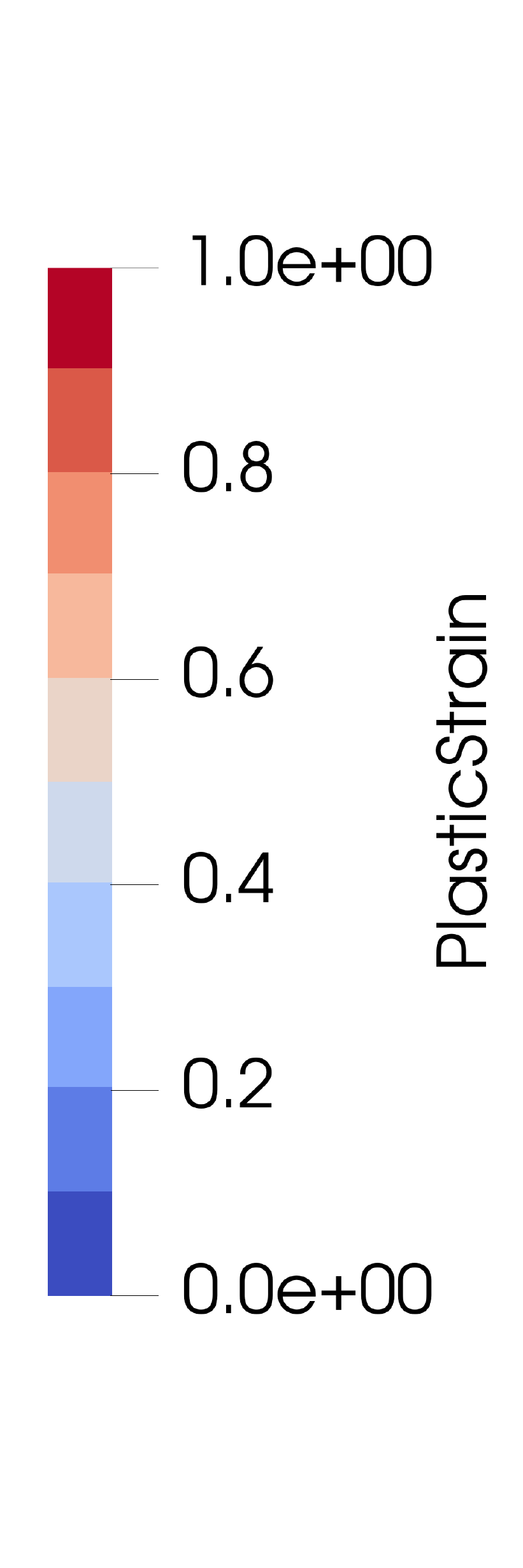}		
	\caption{Pillar final deformation for  the self interaction model. Accumulated plastic strain  corresponding to  $\epsilon^{engn}= 18\%$   for two  choices of initial   dislocation densities: case (i)   (left) and  case (ii)  (right).} \label{fig:PS}
\end{figure}
Firstly, we investigate the compression of pillars under the assumption of self-interaction to analyze the disparities in the mechanical response between the two cases defined in subsection \ref{Sat2D} with two choices of the initial  distribution of the dislocation densities corresponding to small and large specimen.

\textit{Final shape.} As shown in Fig. \ref{fig:PS}, the micro-pillars exhibit qualitatively different deformation processes. In case (i), plastic deformation is localized, taking the form of a thin shear band that accumulates almost all the deformation (the  accumulated plastic strain exceeds 100\%), effectively separating two rigid regions. Conversely, in case (ii), the response is typical of conventional bulk materials, where the deformation is more  homogeneous, leading to a barrel-shaped pillar. 
\begin{figure}
	\center
	\makebox[\textwidth][c]{\includegraphics[width=1.\textwidth]{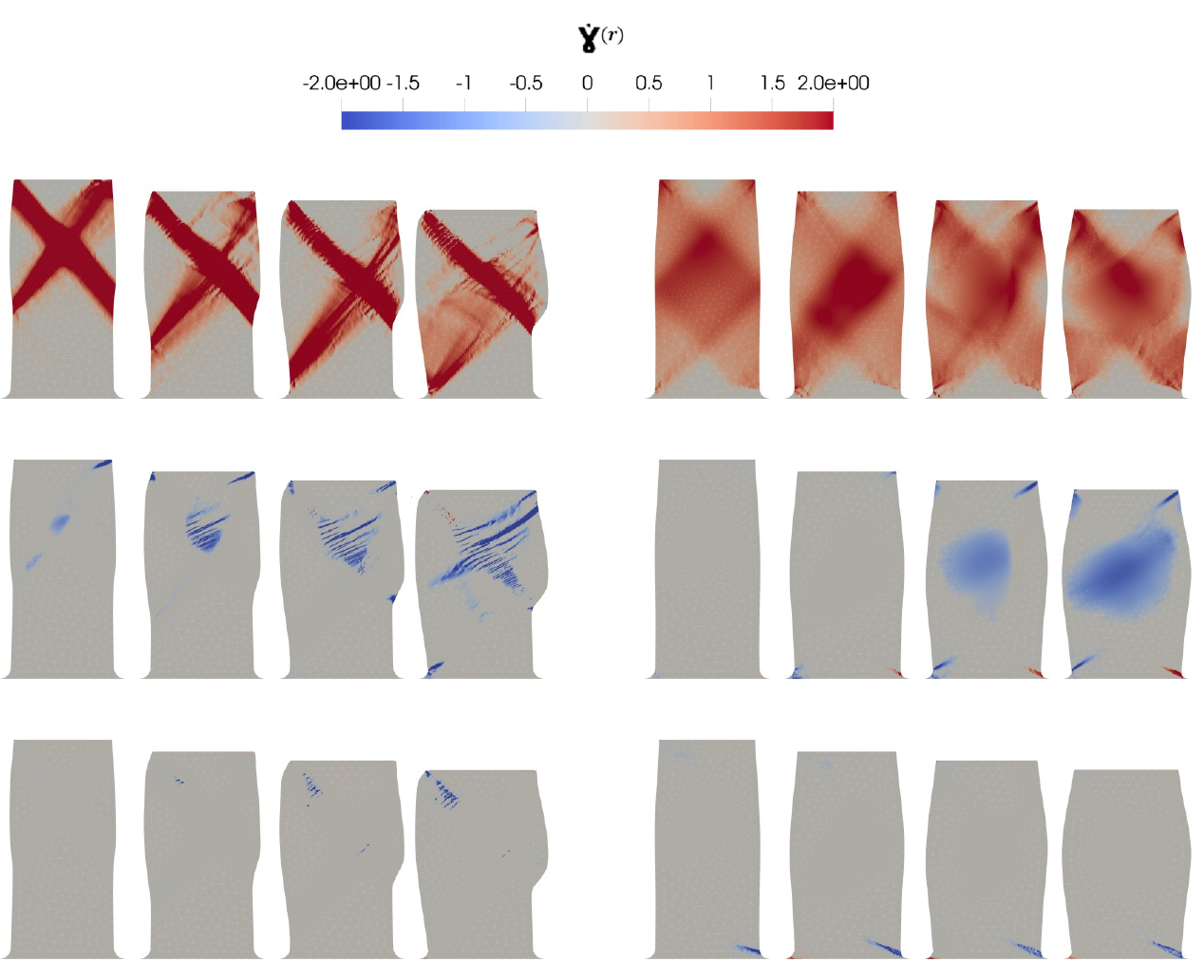}}
	\caption{Pillar deformation for the self interaction model. The slip rates $\dot{\gamma_{1}}$ (up),$\dot{\gamma_{2}}$ (middle) and $\dot{\gamma_{3}}$ (bottom) distribution (in sec$^{-1}$) at different levels of engineering deformations(5$\%$, 10$\%$, 14$\%$ and 18$\%$) for two  choices of initial   dislocation  densities:   case (i)    (left) and  case (ii) (right).}
	\label{fig:Slipself}
\end{figure}

\textit{Slip rates.} Fig. \ref{fig:Slipself} shows the spatial distribution in  the micro-pillar of the slip rates corresponding to the 3  slip systems at different levels of engineering  deformation.  Slip system 1 remains active throughout the entire deformation process, irrespective  the case considered. The activity of system 1 concentrates in the middle of case (ii) (larger specimen), while in case (i), it localizes within the shear band. On the other hand, system 2 is active only for case (ii)  and at higher engineering deformation levels (beyond 10\%).  In the case (i) the activity of system 2 is somewhat limited and confined to a narrow bands. Notably, system 3 remained inactive across all cases.
\begin{figure}
	\center
	\makebox[\textwidth][c]{\includegraphics[width=1.\textwidth]{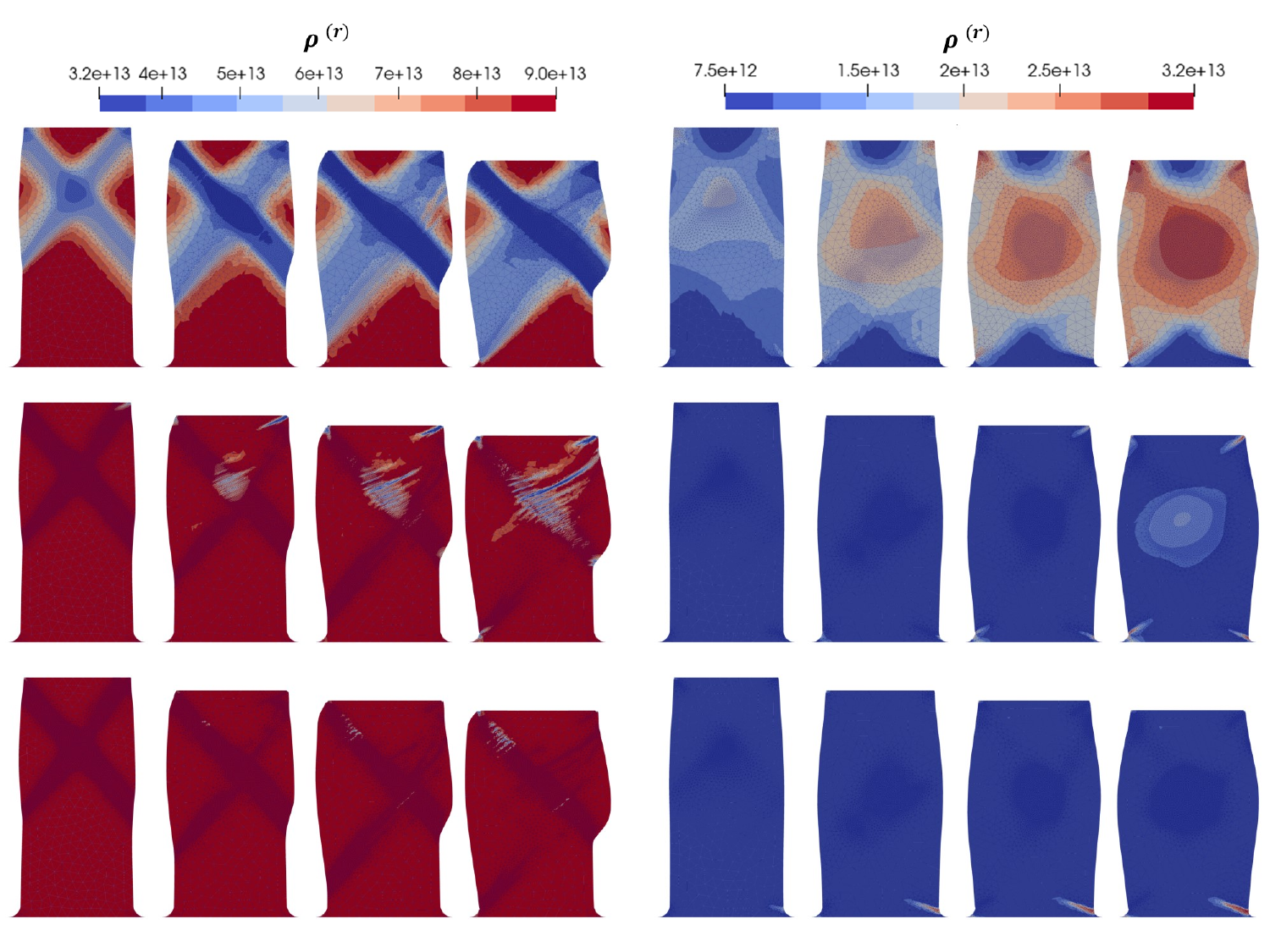}}
	\caption{Pillar deformation for  the self interaction model. Dislocation densities $\rho^{1}$ (up), $\rho^{2}$ (middle) and $\rho^{3}$ (bottom) distributions (in m$^{-2}$) at different levels of engineering deformations (5$\%$, 10$\%$, 14$\%$ and 18$\%$) for two  choices of initial   dislocation  densities:   case (i)    (left) and  case (ii)  (right). } 
	\label{fig:DDself}
\end{figure}

\textit{Evolution of dislocation densities.} We show the dislocation density evolution,  in Fig. \ref{fig:DDself}. For slip systems 2 and 3, which have small slip rates, we observe no significant variation. The dislocation density associated with slip system 1 decreases as deformation progresses in case (i) (small sample), while it increases in case (ii) (larger sample), as expected from the stability analysis described in subsection \ref{Sat2D}.
As it follows from Figure \ref{DD_tau_self} right, in case (i) (smaller pillar), we expect  a softening deformation process, whereas in case (ii) (larger pillar), a hardening deformation process occurs.  

\textit{Saturation  of dislocation densities.}  The saturation of the computed dislocation densities starts at a high level of deformation and affects only system 1. This is evident for case (i)  at 12.5\% of  engineering deformation and for case (ii) at 18\%. The delayed saturation of dislocation densities in case (ii)  is attributed to the substantial difference in slip values, $\gamma^1$, between the two cases at an equivalent total deformation. 
 At 12.5\% of  engineering  strain, the mean slip value $\gamma^1$ was around 90\% in the shear band for case (i), while for case (ii), it did not even reach 40\%. These results confirm that saturation occurs for larger values of slip (more than 50\%), as pointed out by the previous stability analysis.

\textit{Shear bands.} Altogether, this means that when dealing with softening processes (as in the case (i)), plastic deformation tends to favor the propagation along a specific slip system, leading to one or two shear bands. When hardening is present (as in the case (ii)), shear bands cannot develop significantly, as the plastic yield limit increases with the slip accumulated in the shear band. This is why we observe a multitude of small shear bands, and an  "isotropization" is induced by the presence of multi-slips in different regions of the pillar, resulting in an overall homogeneous deformation. 

\begin{figure}
	\center
	\includegraphics[width=2.75cm,keepaspectratio]{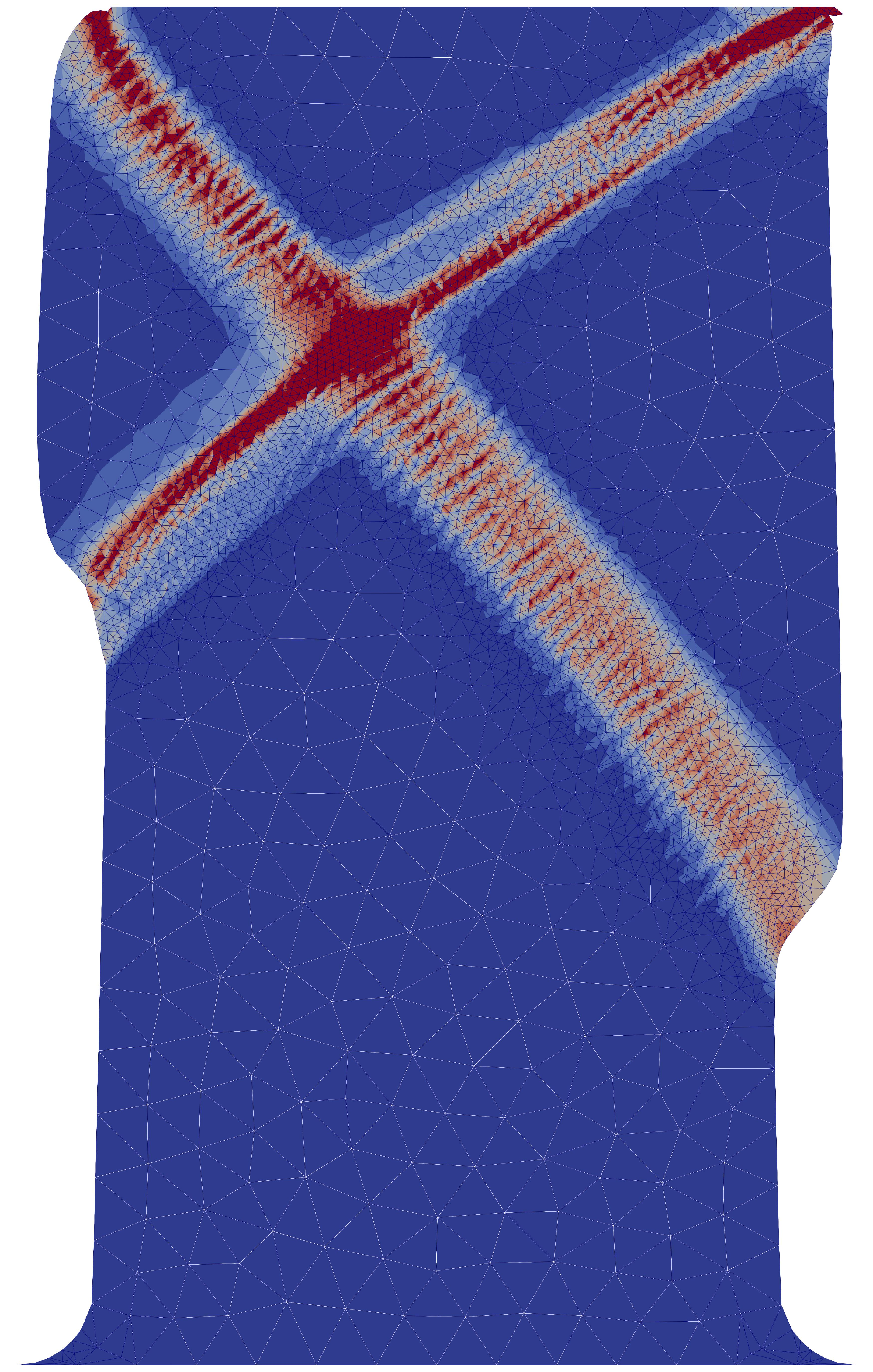}
	\includegraphics[width=2.75cm,keepaspectratio]{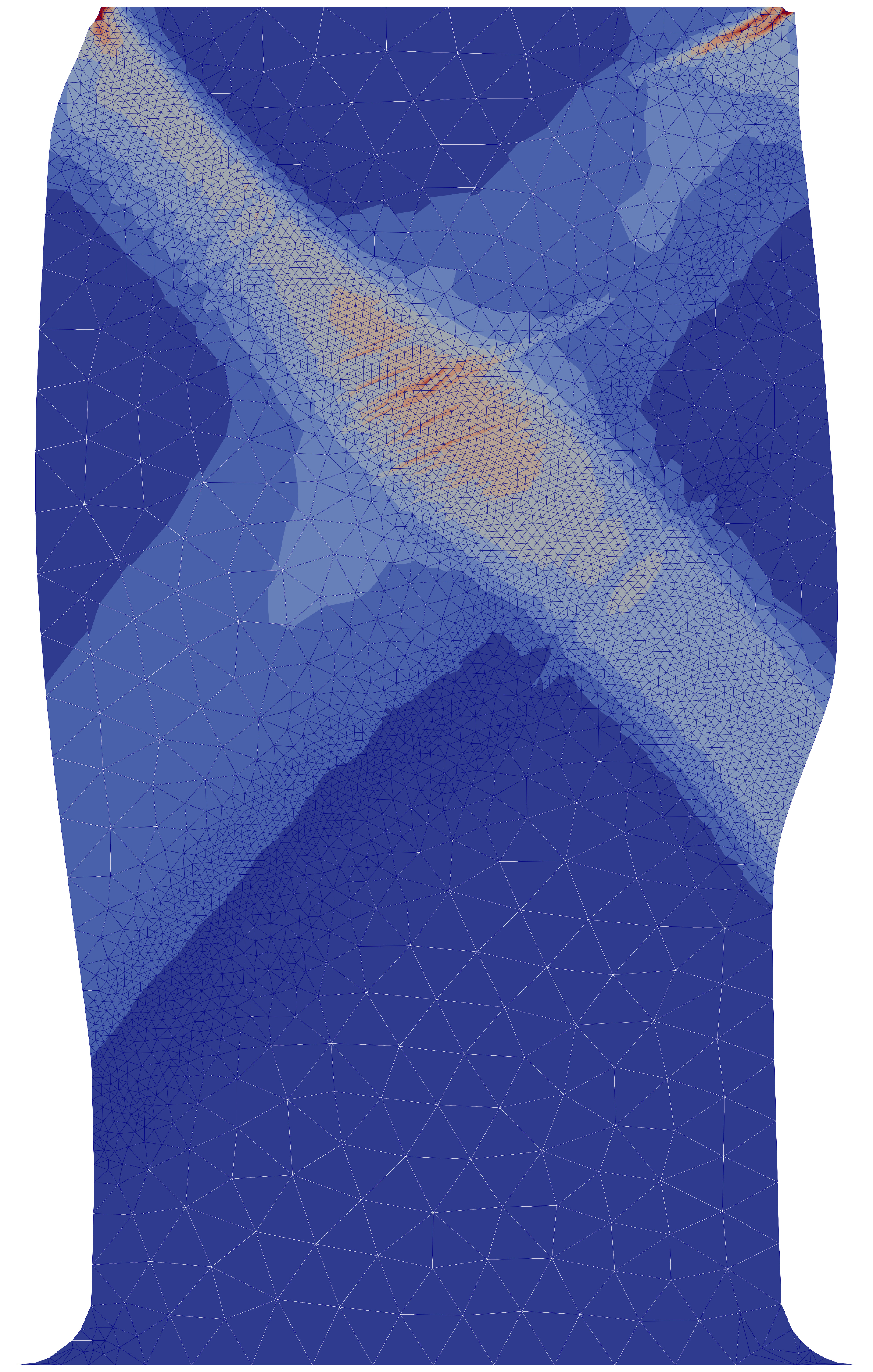}
	\includegraphics[width=2.75cm,keepaspectratio]{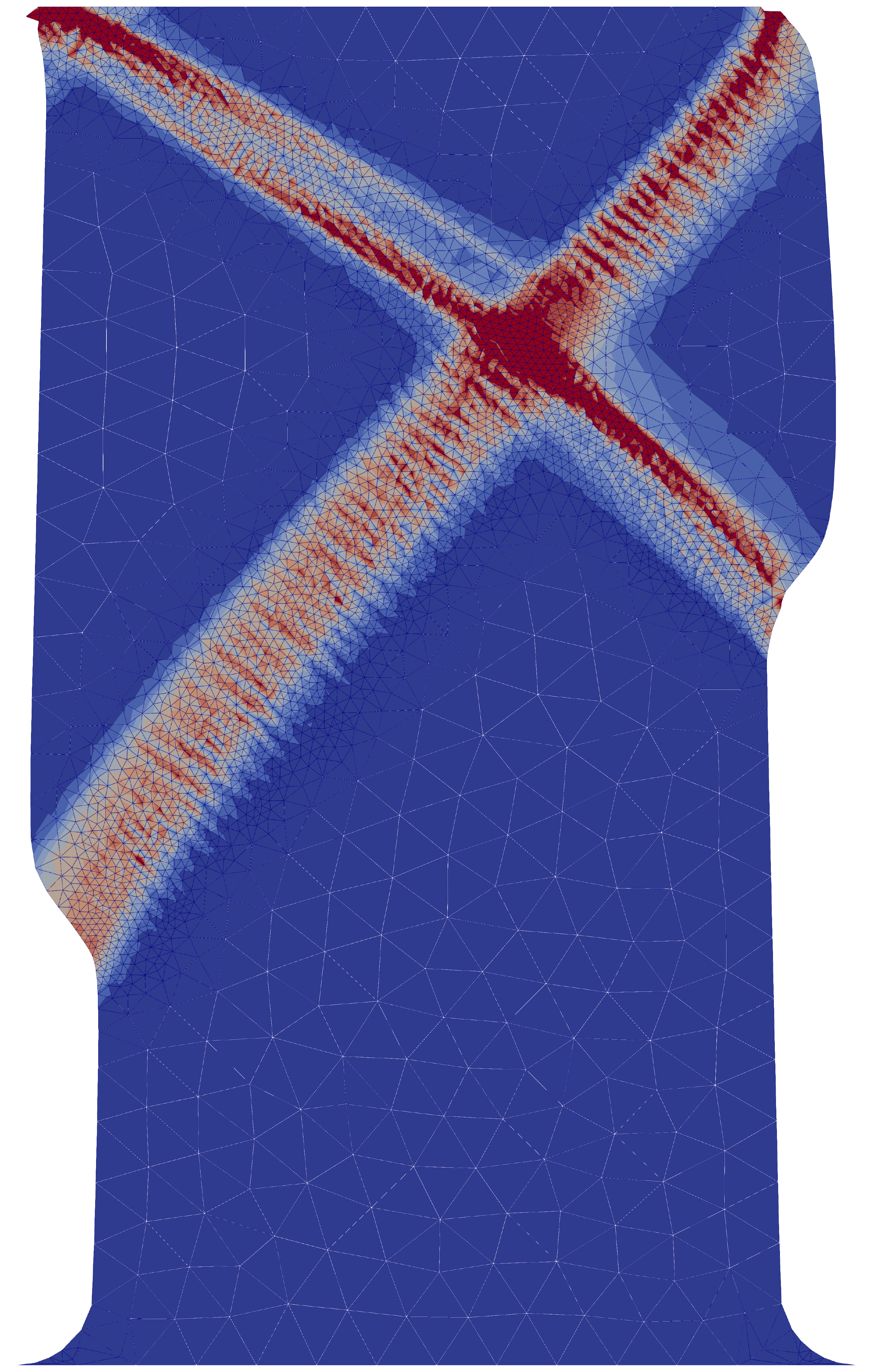}
	\includegraphics[width=2.75cm,keepaspectratio]{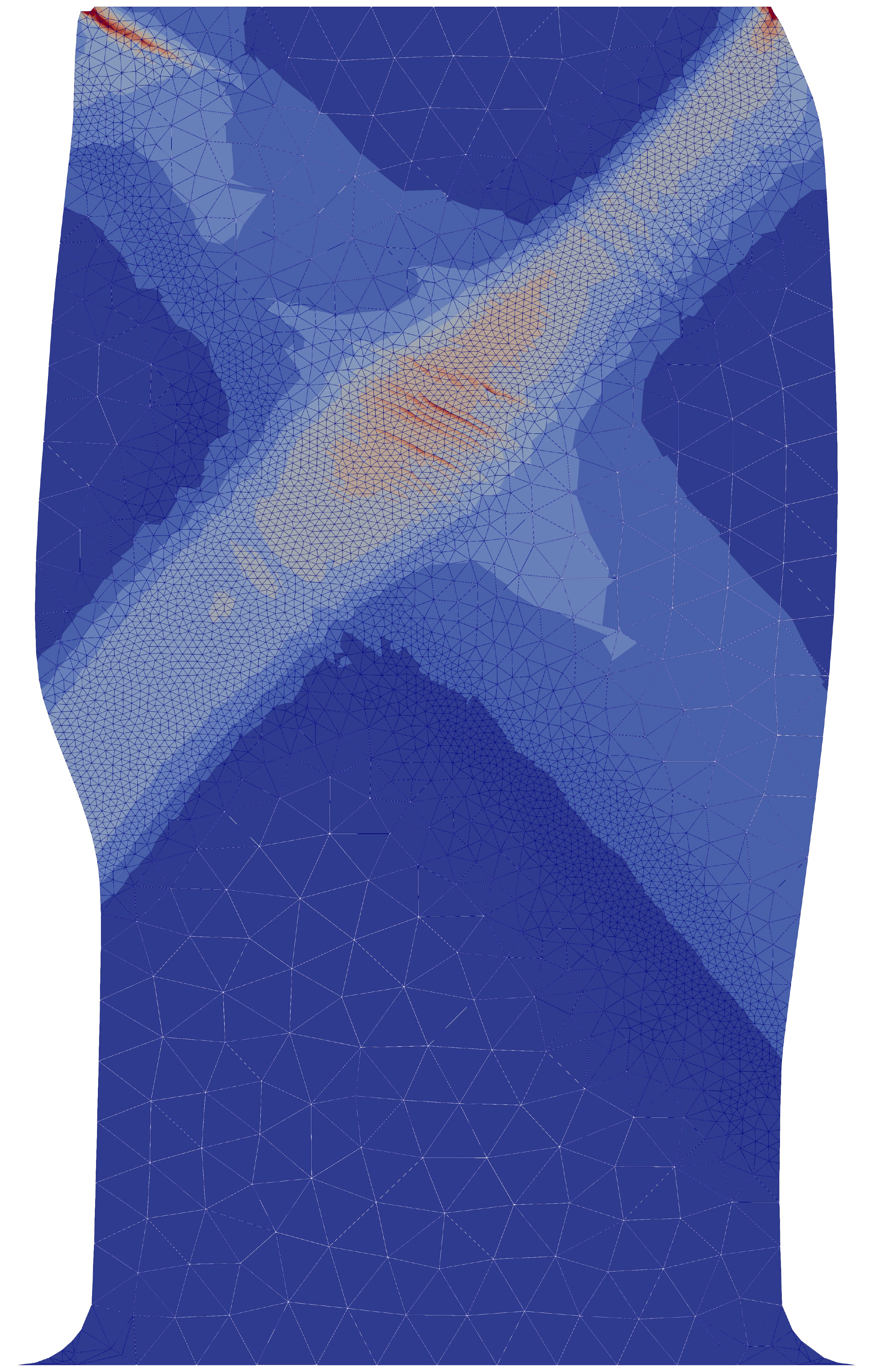}
	\includegraphics[width=1.25cm,keepaspectratio]{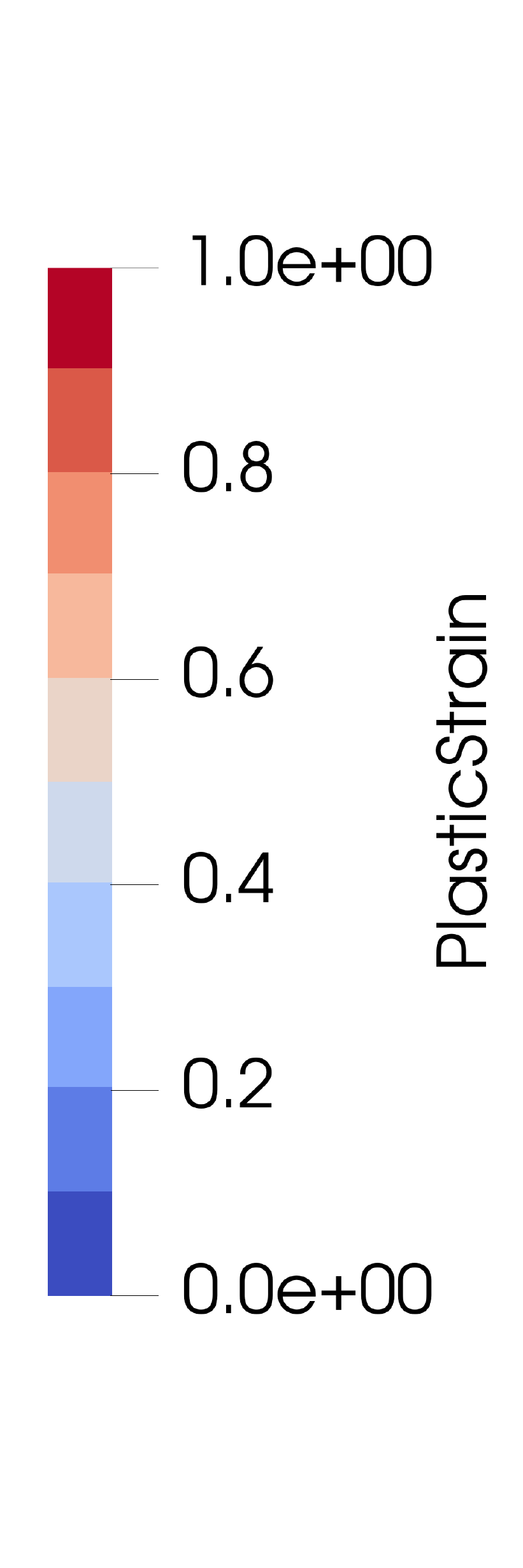}
	\caption{Pillar final deformation for  the self interaction model in case (i). The distribution of cumulated plastic strain at  $\eps^{engn}=11\%$  for different initial orientations of the crystal $\theta^{0}=35^\circ$;  $65^\circ$;  $145^\circ$ and  $115^\circ$.}\label{rotdepisur2}
\end{figure}

\textit{Role played by the initial lattice orientation $\theta^0$.} 
The simulation results presented earlier are based on an initial crystal lattice orientation of $\theta^0 = 65^{\circ}$, which serves as the reference configuration. Our goal is to investigate how the activity of slip systems evolves with different initial orientations. To explore the impact of the initial lattice orientation on localized plastic deformation (case (i)), we conducted additional simulations under the same conditions but with three distinct initial orientations: $\theta^0 = 35^{\circ}$, $\theta^0 = 145^{\circ}$, and $\theta^0 = 115^{\circ}$, depicted in Fig. \ref{rotdepisur2}. For $\theta^0 = 35^{\circ}$, as shown in Fig. \ref{rotdepisur2}(a), the simulation reveals a multislip scenario, characterized by two localized shear bands instead of just one. Interestingly, these shear bands exhibit divergent orientations, appearing to point in opposite directions. Unlike the reference configuration, system 1 is no longer active, allowing systems 2 and 3 to accommodate the deformation.
Assuming that we initially rotate the crystal by $\frac{\pi}{2}$ from the reference configuration, no change in the deformation pattern is observed; deformation localizes exactly as it does in the reference configuration. This consistency persists even when the reference configuration is set at an angle of $35^\circ$. Alternatively, when the initial rotation angle is set as $\pi - \theta$, an intriguing shift in behavior is observed. The shearing direction is inverted relative to the shearing direction associated with the $\theta$ orientation, exhibiting mirror symmetry, as shown in Fig. \ref{rotdepisur2}. Irrespective of the specific initial angle $\theta^0$ chosen, a consistent observation emerges: stable stationary shear bands form with an absolute angle between $45^\circ$ and $55^\circ$.
\subsection{Cross-interaction simulations}

In this section, we explore the features captured by cross-interaction calculations and compare them to the results from self-interaction. .                             

\textit{Final shape.} Notably, for case (ii), the deformation process (i.e., the final shape) remains qualitatively unchanged, consistently exhibiting a hardening process characterized by diffusive overall deformation. However, a distinctive observation emerges in the case (i). Here,  the cross-interaction case reveals aspects related to dislocation interactions between different systems that are not captured by the self-interaction model. The difference between this case and self-computation becomes evident in this context.

\begin{figure}
	\center
	\makebox[\textwidth][c]{\includegraphics[width=1.0\textwidth]{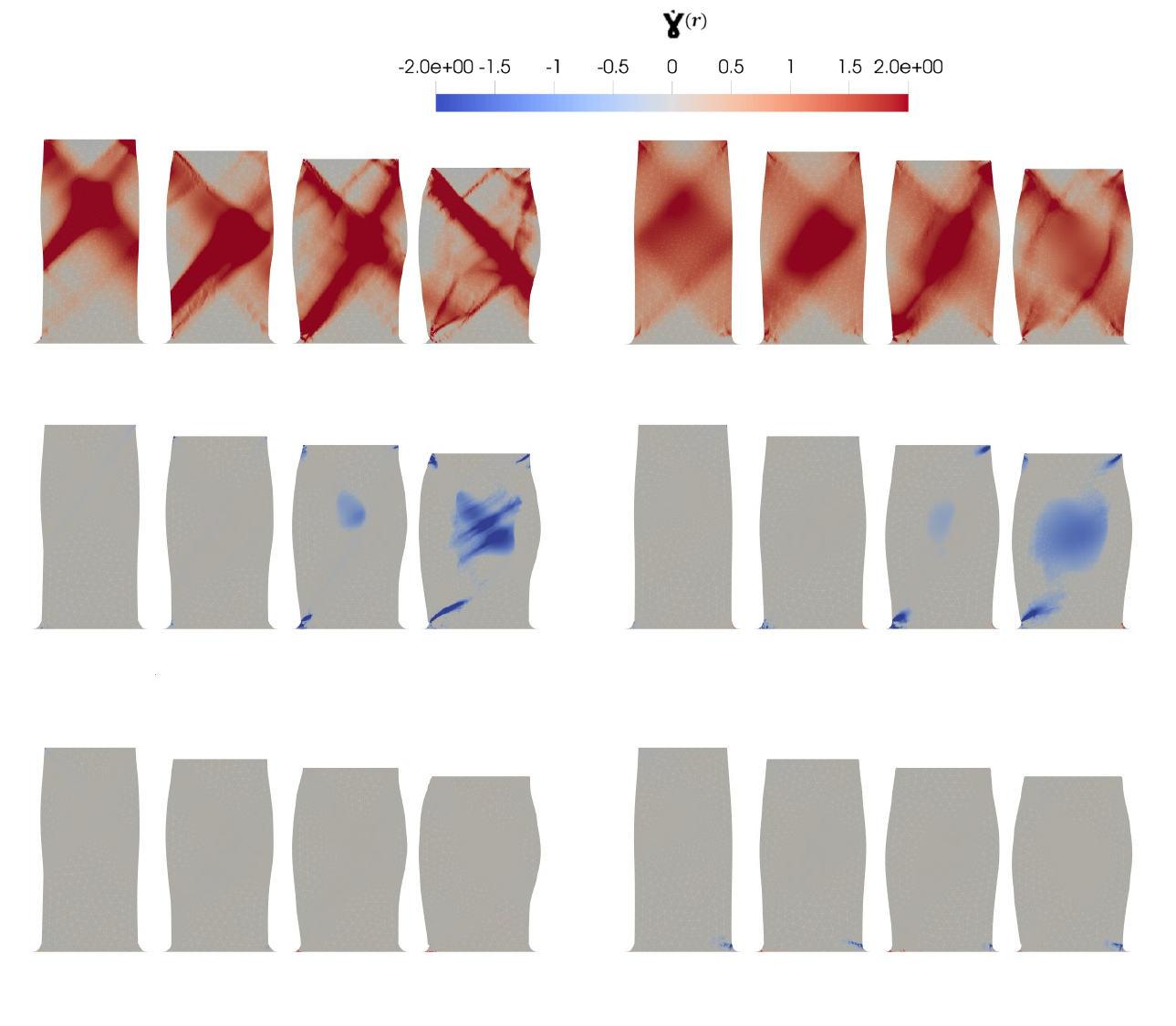}}
	\caption {Simulation with the cross-interaction (KM)  model. The slip rates $\dot{\gamma_{1}}$ (up),$\dot{\gamma_{2}}$ (middle) and $\dot{\gamma_{3}}$ (bottom) distribution (in sec$^{-1}$) at different levels of engineering deformations (5$\%$, 10$\%$, 14$\%$ and 18$\%$). Left: case (i) (associated to $D=1 \mu$m). Right:   case (ii) (associated to $D=20 \mu$m).} 
	\label{fig:SlipUC1}
\end{figure}

\textit{Slip rates.}  
\textcolor{black}{
Examining Fig.~\ref{fig:SlipUC1}, which depicts the spatial slip rate distributions, we observe similarities between the cross and self-interaction computations. In both samples, slip system 3 remains inactive during deformation, while slip system 1 is consistently active. Significantly, the activation of system 2—occurring at a later deformation stage (beyond 14\%)—partly obstructs the activity of system 1, constituting a clear divergence between the cross and self-interaction computations. In the cross-interaction computation, the slip motion orientation is no longer dictated solely by system 1, in contrast to the self-interaction model. Instead, the slip direction is influenced by both systems 1 and 2, with a bias toward system 1.
}

\begin{figure}
	\center
	\makebox[\textwidth][c]{\includegraphics[width=1.\textwidth]{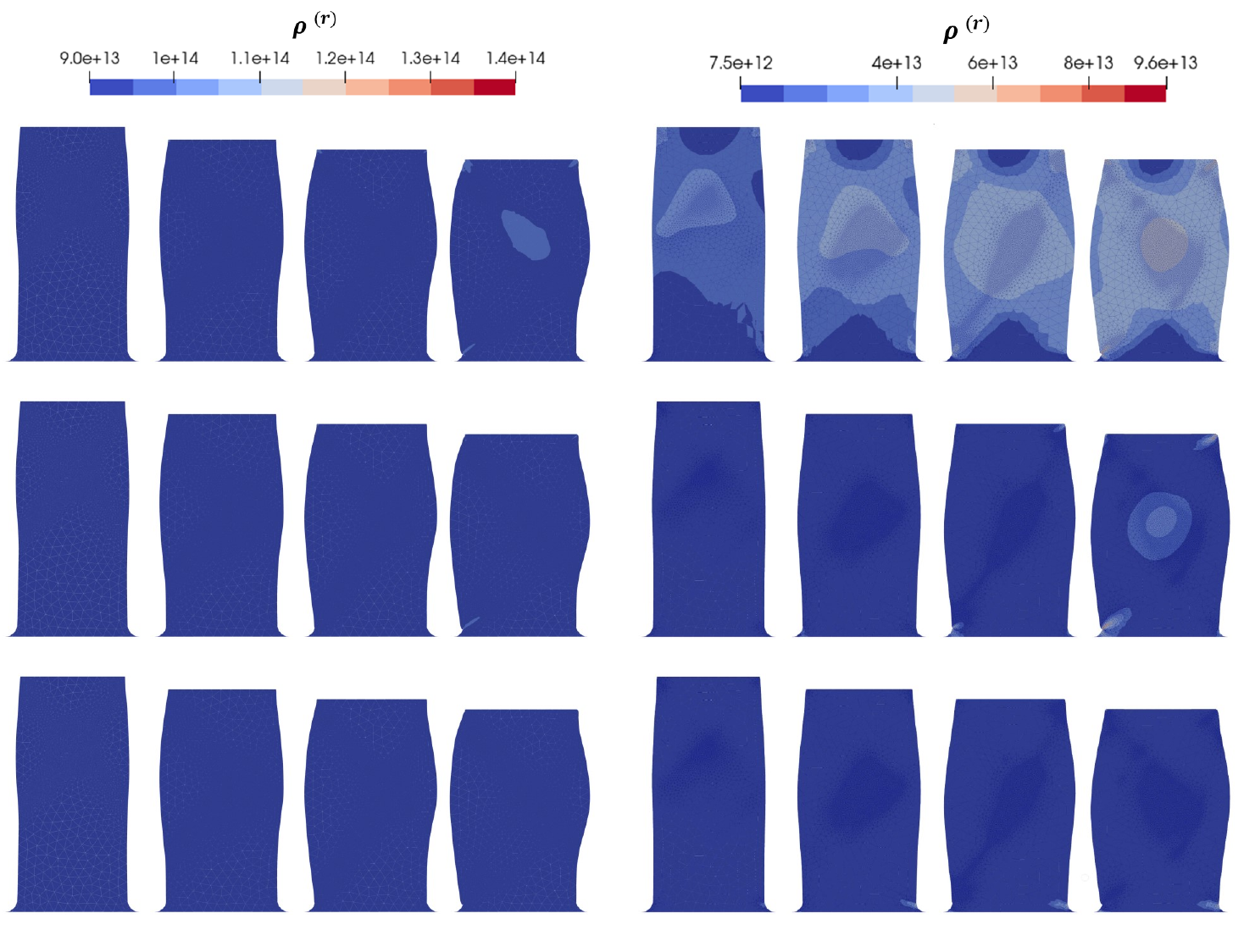}} 
	\caption {Simulation with the cross-interaction (KM) model. Dislocation densities $\rho^{1}$ (up), $\rho^{2}$ (middle) and $\rho^{3}$ (bottom) distributions (in m$^{-2}$) at different levels of engineering deformations (5$\%$, 10$\%$, 14$\%$ and 18$\%$).  Left: case (i) (associated to $D=1 \mu$m). Right:   case (ii) (associated to $D=20 \mu$m).} 
	\label{fig:DDUC1}
\end{figure}

\textit{Evolution of dislocation densities.}  According to the above observation concerning the active slip systems we conclude that this situation corresponds to case (a) of the stability analysis of section \ref{Att2D} and depicted in Figure \ref{fig:DD_tau_UC1}. The evolution of spatial distributions of dislocation densities in the cross  interaction computation are  shown in Fig. \ref{fig:DDUC1}.

\textcolor{black}{\textit{Saturation of dislocation densities.} In case (ii), the stability analysis predicts an increase in the dislocation density of system 1, which is associated with a hardening process, unlike the plateau observed in case (i). This leads to a change in the deformation mode of the pillar: shear bands cannot form, and instead, diffuse deformation is observed. This implies that the slip strain is insufficient for system 1 (and eventually system 2) to reach the saturation dislocation density and the corresponding saturation yield stress $\tau_{c,\text{sat}}^{1,(ii)}$.
}

\section{Conclusions}
\label{sec:conclusions}
Dislocation-density based crystal plasticity models are introduced to account for the microstructral changes throughout the deformation process, enabling more accurate predictions of deformation process compared to slip-system resistance-based plasticity models.  In this work, we presented a stability analysis of some established dislocation density-based models, including the Kocks and Mecking (KM) model and its variants, aiming to identify conditions for stationary states in active slip systems and evaluate their linear stability. Our analysis can be generalized to any type of dislocation density model, providing a broader framework for understanding the stability of such systems.    Following this analysis, we find the presence of saturated dislocation densities and the essential role of initial dislocation density {\color{black} in distinguishing between hardening and softening responses}. Given that the initial dislocation density could be related to the sample's size or preparation process, size-effects can potentially be captured.

Finally, we conducted numerical simulations of micro-pillar compression  using an Eulerian crystal plasticity approach to address the behavior of materials at micro-scale dimensions. Our findings indicate that microstructural evolution in small-scale materials can be effectively modeled using dislocation-density based CP models, providing valuable insights for the design of miniaturized mechanical devices and advanced materials in the rapidly evolving field of nanotechnology.
\textcolor{black}{
To be more precise, the impact of initial dislocation density on the mechanical response in monocrystals is significant, as shown in several studies 
\cite{Norfleet2008-oc, BYER20133808, BEI2007397, Rinaldi2012}.
These studies emphasize that variations in microstructure, particularly initial dislocation density, can arise due to differences in sample size and/or fabrication methods.}

\textcolor{black}{For instance, the situation when the fabrication method change or alters the microstructure, micro-pillars milled from bulk single crystals using a focused ion beam (FIB), this process shapes the pillars by bombarding the target's surface with gallium-accelerated heavy ions, modifying the microstructure of specimens with a large surface-to-volume ratio. This modification includes introducing an additional dislocation network, decorating dislocations with gallium, or forming gallium precipitates~\cite{Borasi2024}. The FIB preparation effect on mechanical response is strong for ordered bulk material (low dislocation density) and becomes more pronounced with decreasing sample size. As a result, the initial dislocation density increases with decreasing sample dimensions \cite{CUI2014279}. This inverse relationship between sample size and initial dislocation density is referred to in the literature as a "size-effect", which is notably absent in conventional bulk plasticity. Furthermore, it has been reported that regardless of sample preparation, two (Cu) nano specimens with the same initial microstructure exhibit identical mechanical responses, even if their sizes differ \cite{Jennings2010-yg}. This observation suggests that the initial microstructure may have a more significant impact than size, leading to a preference for discussing the "initial microstructure effect" rather than the "size-effect".}


\end{document}